\documentclass{LMCS}

\def\dOi{11(3:4)2015}
\lmcsheading%
{\dOi}
{1--27}
{}
{}
{Feb.~14, 2014}
{Aug.~18, 2015}
{}

\ACMCCS{[{\bf Theory of computation}]: Logic---Proof Theory\,/\,Modal and temporal logics.
}

\usepackage{t1enc}
\usepackage[utf8]{inputenc}

\usepackage{amsmath}
\usepackage{amsfonts}
\usepackage{amssymb}
\usepackage{mathtools}

\usepackage{ae,aecompl}

\usepackage{graphicx}
\usepackage{stmaryrd}
\usepackage{xcolor}
\usepackage{tikz} 
\usetikzlibrary{arrows,shapes,calc}
\usepackage{bussproofs}
\usepackage{wrapfig}
\usepackage{hyperref}

\usepackage{roadlogic}
\usepackage{proofs}

\title[Proof Theory of a Multi-Lane Spatial Logic]{Proof Theory of a Multi-Lane Spatial Logic}

\author{Sven Linker}
\address{Carl von Ossietzky Universität Oldenburg\\
Department für Informatik\\
26111 Oldenburg\\
Germany}
\email{\{linker,hilscher\}@informatik.uni-oldenburg.de} 

\author{Martin Hilscher} 
\address{\vspace{-18 pt}}

\begin{document}

\keywords{Spatial logic,
Undecidability,
Labelled natural deduction.}

\titlecomment{{\lsuper*}This research
    was partially supported by the German Research Council (DFG) in
    the Transregional Collaborative Research Center SFB/TR 14 AVACS.
This paper is the extended and slightly revised version of 
our publication in the 10th International 
Colloquium on Theoretical Aspects of Computing (ICTAC) in 2013 \cite{LH2013}.}

\EnableBpAbbreviations
\begin{abstract}
We extend the Multi-lane Spatial Logic MLSL, introduced in
previous work
for proving the safety (collision freedom) of
traffic maneuvers on a multi-lane highway, by length measurement and
dynamic modalities. We investigate the proof theory of this extension,
called EMLSL.
To this end, we prove the undecidability of EMLSL but
nevertheless present a sound proof system which allows for reasoning
about the safety of traffic situations. 
We illustrate the latter by giving a formal proof for the \emph{reservation lemma} we could
only prove informally before. Furthermore we prove a basic theorem  
showing that the length measurement is independent from the number of 
lanes on the highway. 
\end{abstract}

\maketitle

\section{Introduction}
\label{sec:intro}

In our previous work \cite{HLOR2011} we proposed a multi-dimensional
spatial logic MLSL inspired by Moszkowski's interval temporal logic (ITL)
\cite{Mos1985}, Zhou, Hoare and Ravn's Duration
Calculus (DC) \cite{CHR1991} and Sch\"afer's Shape Calculus
\cite{Sch2005} for formulating the  
purely spatial aspects of safety of traffic maneuvers on
highways. In MLSL we modeled the highway as one continuous dimension,
i.e., in the direction along the lanes and one discrete dimension, the
different lanes. We illustrated MLSL's usefulness by 
proving  safety of two variants of lane change maneuvers on
highways.
The safety proof establishes that the braking distances of no two
cars intersecting is an inductive
invariant of a transition system capturing the dynamics of cars and
controllers.

In this paper we introduce EMLSL which extends MLSL by length
measurement and dynamic modalities. In comparison to MLSL, where we
are only able to reason about  qualitative spatial properties, i.e., 
topological relations between cars, EMLSL also allows for quantitative reasoning,
e.g., on braking distances. To further the practicality of EMLSL, we define
a proof system  based on ideas of 
Basin et al.~\cite{BMV1998}, who presented
systems of  labelled
natural deduction for a vast class of typical modal logics.  
Rasmussen~\cite{Ras2001} 
refined their work  to interval
logics with binary chopping modalities. Since EMLSL incorporates both
unary as well as chopping modalities, our proof system is strongly related to 
both approaches. 

Besides providing a higher expressiveness,
extending MLSL enables us to formulate and prove the invariance of the
spatial safety property \emph{inside} EMLSL and its deductive proof 
system.
We demonstrate this by conducting a formal proof of the so called
\emph{reservation lemma} \cite{HLOR2011}, which informally states that
no car changes lanes without having set the turn signal
beforehand. 

Further on, 
we show undecidability 
of a subset of EMLSL.  
We adapt the
proof of Zhou et al.~\cite{CHS1993} for DC and reduce the halting
problem of two
counter machines to satisfiability of EMLSL formulas. Due to the
restricted set of predicates EMLSL provides, this is non-trivial.

The \emph{contributions} of this paper are as follows:
\begin{itemize}
  \item we extend MLSL with lengths measurements and
    dynamic modalities (Sect.~\ref{sec:mlsl});
  \item we show the spatial fragment of EMLSL to be undecidable
    (Sect. \ref{sec:undec});
    \item we present a suited proof system  
and derive 
the reservation lemma (Sect.~\ref{sec:lnd}).
\end{itemize}

The \emph{differences} to our publication in the proceedings 
of the 10th International Colloquium on Theoretical Aspects of 
Computing (ICTAC) in 2013 \cite{LH2013} are:

\begin{itemize}
\item we include the proofs for the preservation of sanity conditions of the spatial
situations along the transitions (Sect.~\ref{sec:mlsl}), the undecidability result (Sect.~\ref{sec:undec}) and the soundness 
      of the proof system (Sect.~\ref{sec:lnd});
\item we show an additional formal proof within the proof system for a theorem showing
the independence of length measurement from the width (i.e., the number of lanes
currently perceivable by a car; see Lemma~\ref{lem:length_and_width}). This proof is straightforwardly adaptable to a
proof for the reversed situation, i.e., the independence of width measurement from
the extension (the part of the highway currently perceived by a car in driving
direction);
\item we added means of \emph{moving} the part of the highway perceived by a car 
along the passing of time in Sect.~\ref{sec:mlsl}. This addition has also impact on the form of the labelling
algebra of the proof system in Sect.~\ref{sec:lnd}.
\end{itemize}


\section{Extended MLSL Syntax and Semantics}
\label{sec:mlsl}

The purpose of EMLSL is to reason about highway situations. To this end, 
we  first present the formal model of a \emph{traffic snapshot} capturing
the position
and speed of every car on the highway at a given point in time. In
addition a traffic snapshot comprises the lane a given car is driving on, which
we call a \emph{reservation}. Every car usually holds one
reservation, i.e., drives on one lane, but may, during lane change
maneuvers, hold up to two reservations on adjacent lanes. Furthermore,
we capture the indication that a given car wants to change to a
adjacent lane by the notion of a \emph{claim} which is an abstraction of
setting the turn signal. Every car may only hold claims while
not  
engaged in a lane change.

\begin{figure}[htb]
  \centering
  \includegraphics{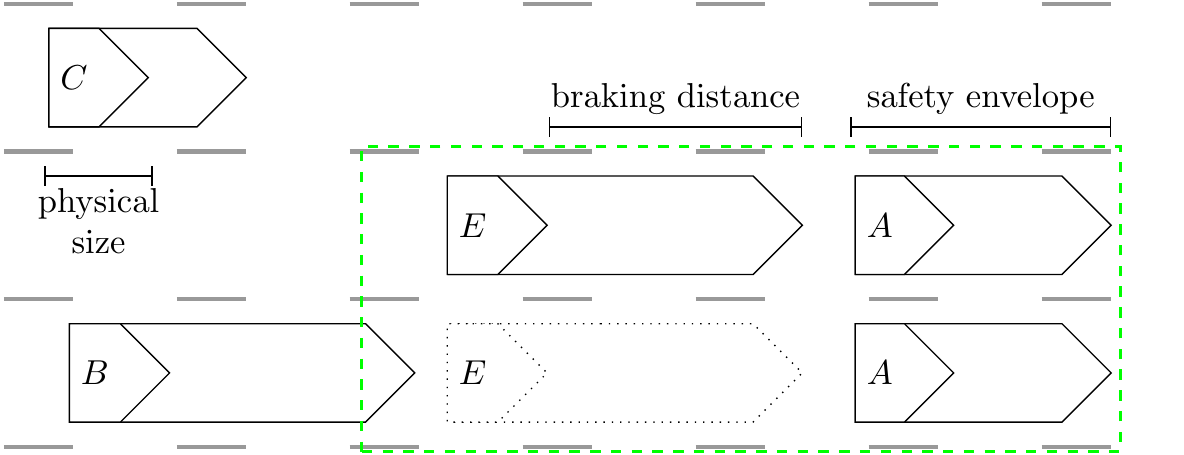}
  \caption{Situation on a Motorway at a Single Point in Time}
  \label{fig:example_situation}
\end{figure}

Intuitively, traffic snapshots shall formalize situations as depicted
in Fig.~\ref{fig:example_situation}. Each car drives at a certain
horizontal position and reserves one or at most two lanes.
The car \(E\) is currently claiming the lower lane, depicted
by the dotted polygon. For a car, we subsume its physical size and its 
braking distance, i.e., the distance it needs to come to
a safe standstill at its current speed, under its \emph{safety envelope}.
As an abstraction of sensor limitations, we assume each car
to observe only a finite part of the road, called the \emph{view}
of the car. 
The dashed rectangle indicates a possible view of the car \(E\).

To formally define a traffic snapshot, we 
assume a countably infinite set of globally unique \emph{car identifiers} 
\(\ID\) and
an arbitrary but fixed set of lanes \(\Lanes =  \{0, \ldots , N\}\), for
some \(N \geq 1\).
Throughout this paper we will furthermore make use of the notation
\(\Powerset{X}\) for the powerset of \(X\), and the override
notation \(\oplus\) from Z for function updates \cite{WD96},
i.e., \(f\oplus \{x\mapsto y\}(z) = y\ \text{if}\ x=z\ \text{and}\ f(z)\
\text{otherwise}\).

\begin{defi}[\TSText]
\label{def:snapshots}
A \emph{traffic snapshot} 
\(\RoadDef\) 
is defined by the functions 
\begin{itemize}
  \item $\laneReservation:\ID\rightarrow\Powerset{\Lanes}$ such that
    {\(\laneReservation(C)\) is the set of lanes the car \(C\) reserves,}
  \item $\laneClaim: \ID\rightarrow \Powerset{\Lanes}$ such that 
  {\(\laneClaim(C)\) is the set of lanes the car \(C\) claims,}
  \item $pos : \ID\rightarrow\R$ such that 
  {$\pos(C)$ is the position of the car $C$ along the lanes,}
  \item $\spd: \ID\rightarrow\R$ such that 
  {$\spd(C)$ is the current speed of the car $C$,}
  \item $\acc: \ID\rightarrow\R$ such that 
  {$\acc(C)$ is the current acceleration of the car $C$.}
\end{itemize}
Furthermore, we require the following \emph{sanity conditions} to hold for all 
\(C\in\ID\). 
\begin{enumerate}
\item \(\laneReservation(C) \cap \laneClaim(C) = \emptyset\) \label{prop:recl_mutex}
\item \(1 \leq |\laneReservation(C) | \leq 2\) \label{prop:num_res}
\item \(0 \leq |\laneClaim(C)|\leq 1\) \label{prop:num_clm}
\item \(1 \leq |\laneReservation(C)| + |\laneClaim(C)| \leq 2\) \label{prop:max_lanes}
\item \(\laneClaim(C) \neq \emptyset \text{ implies } \exists n \in \Lanes \qsep \laneReservation(C) \cup \laneClaim(C) = \{n, n+1\}\) \label{prop:clm_adj_res}
\item \(|\laneReservation(C)| = 2 \) or \(|\laneClaim(C)| = 1\) holds only for finitely many \(C\in \ID\). \label{lem:clm_res_fin} 
\end{enumerate}
We denote the set of all traffic snapshots by \(\Roads\).
\end{defi}

The kinds of transitions are twofold.
First, we have discrete transitions defining the possibilities to create, 
mutate and remove claims and reservations. The other type of transitions 
handles abstractions of the dynamics of cars, i.e., they allow for 
instantaneous changes of accelerations and for the passing of time, during which the cars
move according to a simple model of motion. 
For the results 
presented subsequently, we only
require the changes of positions to be continuous.  
\begin{defi}[Transitions]
\label{def:transitions}
The following \emph{transitions} describe the changes that may occur
at a \tsText{} \(\RoadDef\). 
{
\allowdisplaybreaks
\begin{align}
  \Road \transition{\claimLane{C}{n}} & \Road^{\prime} &\Leftrightarrow\nonumber
                    &&&\,\Road^\prime = (\laneReservation, \laneClaim^\prime, \pos, \spd, \acc)\nonumber\\
                    &&&&\wedge&\, |\laneClaim (C)| = 0
                    \wedge \, |\laneReservation (C)| = 1\nonumber\\
                    &&&&\wedge&\, \laneReservation(C)\cap \{n+1,n-1\}\not=\emptyset\nonumber\\
                    &&&&\wedge&\, \laneClaim^\prime=\laneClaim\oplus\{C\mapsto \{n\}\}\label{claimpt}\\
  \Road \transition{\stopclaimLane{C}} & \Road^{\prime} &\Leftrightarrow\nonumber
                    &&&\, \Road^\prime = (\laneReservation, \laneClaim^\prime, \pos, \spd, \acc)\nonumber\\
                    &&&&\wedge&\, \laneClaim^\prime=\laneClaim\oplus \{C\mapsto\emptyset\}\label{claimmt}\\
  \Road \transition{\moveToLane{C}} & \Road^{\prime} &\Leftrightarrow\nonumber
                    &&&\, \Road^\prime = (\laneReservation^\prime, \laneClaim^\prime, \pos, \spd, \acc)\nonumber\\
                    &&&&\wedge&\, \laneClaim^\prime=\laneClaim\oplus\{C\mapsto\emptyset\}\nonumber\\
                    &&&&\wedge&\, \laneReservation^\prime=\laneReservation\oplus\{C\mapsto \laneReservation(C)\cup \laneClaim(C)\}\label{respt}\\
  \Road \transition{\stopmoveToLane{C}{n}} & \Road^{\prime} &\Leftrightarrow\nonumber
                    &&&\, \Road^\prime = (\laneReservation^\prime, \laneClaim, \pos, \spd, \acc)\nonumber\\
                    &&&&\wedge&\, \laneReservation^\prime=\laneReservation\oplus\{C\mapsto \{n\}\}\nonumber\\
                    &&&&\wedge&\, n\in \laneReservation(C)
                        \wedge \, |\laneReservation(C)|=2\label{resmt}\\
  \Road \transition{t} & \Road^{\prime} &\Leftrightarrow
                    &&&\, \Road^\prime = (\laneReservation, \laneClaim, \pos^\prime, \spd^\prime, \acc)\nonumber\\
                    &&&&\wedge&\, \forall C\in \ID\colon\pos^\prime(C)=\pos(C)+\spd(C)\cdot t+\tfrac{1}{2}\acc(C)\cdot t^2\nonumber\\
                    &&&&\wedge&\, \forall C\in \ID\colon \spd^{\prime}(C)=\spd(C)+\acc(C)\cdot t\label{timet}\\
  \Road \transition{\accelerate{C}{a}}& \Road^{\prime} &\Leftrightarrow\nonumber 
                    &&& \Road^{\prime} = (\laneReservation, \laneClaim, \pos,\spd,\acc^{\prime}) \nonumber\\
                    &&&&\wedge&\, \acc^{\prime}=\acc \oplus\{C\mapsto a\}\label{accelerate} 
\end{align}
}
\end{defi}

We also combine  passing of time and changes of accelerations to  
\emph{evolutions}.
\begin{align*}
  \Road \abstrans{t} \Road^\prime &\metaequiv \Road = \Road_0 \transition{t_0} \Road_1 \transition{\accelerate{C_0}{a_0}} \dots  \transition{t_n} \Road_{2n-1} \transition{\accelerate{C_n}{a_n}} \Road_{2n} = \Road^\prime, 
\end{align*}
where \(t = \sum_{i = 0}^{n} t_i\), \(a_i \in \R\) and \(C_i \in \ID\) 
for all \( 0 \leq i \leq n\).

The transitions  preserve the  
sanity conditions  in Def.~\ref{def:snapshots}.

\begin{lem}[Preservation of Sanity]
  \label{lem:transitions} Let \(\Road\) be a snapshot satisfying the
constraints given in Def.~\ref{def:snapshots}. Then, each structure
\(\Road^\prime\) reachable by a transition is again a traffic snapshot
satisfying Def.~\ref{def:snapshots}. 
\end{lem}
\proof
  We proceed by a case distinction. If the transition leading from
\(\Road\) to \(\Road^\prime\) is the passing of time, or the change of an
acceleration, the constraints are still satisfied in \(\Road^\prime\), 
since they only concern the amount and place of claims and reservations. 

The removal of a claim \(\Road \transition{\stopclaimLane{C}} \Road^\prime\)
sets \(\laneClaim^\prime(C) = \emptyset\). There are two possibilities. 
If \(\laneClaim(C) = \emptyset\), then \(\Road = \Road^\prime\) and hence
satisfies the constraints trivially. Let \(\laneClaim(C) \neq \emptyset\).  
After the transition, constraint 1 holds trivially, constraint 2 is not affected,
constraint 3 holds, as does constraint 4. Constraint 5 holds trivially and
satisfaction of constraint 6 follows since it is satisfied in  \(\Road\) and
we only shrink the number of cars for which there exists a claim. 

Now let \(\Road \transition{\claimLane{C}{n}} \Road^\prime\). Then by definition 
of the transition,  \(\laneReservation(C)\) on \(\Road\) contains 
exactly one element, and \(\laneClaim(C)\) is empty. On \(\Road^\prime\), 
\(\laneClaim^\prime(C)\) contains exactly \(n\). Since 
\(\{n+1,n-1\} \cap \laneReservation(C) \neq \emptyset\), \(n\) cannot be an
element of \(\laneReservation^\prime(C)\). Hence, the constraints 1 to 5 are
satisfied. Since \(\Road\) satisfied constraint 6, and only one car created 
a new claim, \(\Road^\prime\) still satisfies this constraint.

Consider \(\Road \transition{\stopmoveToLane{C}{n}} \Road^\prime\). Since 
\(|\laneReservation(C)| = 2\), constraint 4 ensures that 
\(\laneClaim(C) = \emptyset\), by which constraint 1, 3, 4 and 5 hold in 
\(\Road^\prime\). Constraint 2 holds, since we overwrite \(\laneReservation(C)\)
with \(\{n\}\). Constraint 6 holds by an argument similar to the withdrawal 
of a claim.

Finally, let \(\Road \transition{\moveToLane{C}} \Road^\prime\). Again
we have to consider two cases. First, if \(\laneClaim(C) = \emptyset\), then
\(\Road = \Road^\prime\), and hence the constraints hold. If 
\(\laneClaim(C) \neq \emptyset\), we get by constraint 2 that 
\(\laneClaim(C) = \{n\}\) for some \(n \in \Lanes\). By constraint 4, 
\(|\laneReservation(C)| = 1\), and by constraint 1, we get that after the
transition \(|\laneReservation(C)| = 2\), i.e., constraint 2 holds. Constraint 1 and 5 
hold now trivially. Constraint 3 holds since we reset 
\(\laneClaim^\prime(C) = \emptyset\) and similarly for constraint 4.  The number
of cars with either two reservations or a claim is not changed, hence
constraint 6 holds. \qed

\begin{exa}
We formalize Fig.~\ref{fig:example_situation} as a traffic snapshot
\(\RoadDef\). We will only present the subsets of the functions for
the cars visible in the figure. Assuming that the set of lanes
is \(\Lanes=\{1,2,3\}\), where \(1\) denotes the lower lane and
\(3\) the upper one, the functions defining the 
reservations and claims of \(\Road\) are given by
\begin{align*}
  \laneReservation(A)& = \{1,2\} & \laneReservation(B) &= \{1\} & 
\laneReservation(C)& = \{3\} & \laneReservation(E) &= \{2\} \\
 \laneClaim(A)& = \emptyset & \laneClaim(B) &= \emptyset & 
\laneClaim(C)& = \emptyset & \laneClaim(E) &= \{1\} 
\end{align*}
For the function \(\pos\), we chose arbitrary real values which 
still satisfy the relative positions of the cars in the figure.
Similarly, we instantiate the function \(\spd\) such
that the safety envelopes of the cars could match the figure. For
example, since the safety envelope of \(B\) is larger than
the safety envelope of \(C\),  \(B\) has to drive with 
a higher velocity. For simplicity, we assume that all cars
are driving with constant velocity at the moment, i.e.,
for all cars, the function \(\acc\) returns zero.
\begin{align*}
  \pos(A)& = 28  & \pos(B)& = 3.5 & 
\pos(C) &= 2 & \pos(E) &= 14 \\
  \spd(A)& = 8  & \spd(B) &=14  & 
\spd(C) &= 4  & \spd(E) &= 11  
\end{align*}
This traffic snapshot satisfies the sanity conditions. 

For no traffic snapshot \(\Road^\prime\) and lane \(n\) 
we have \(\Road \transition{\claimLane{E}{n}} \Road^\prime\),
since \(|\laneClaim(E)| \neq 0\). Similarly,
there is no transition \(\Road \transition{\stopmoveToLane{B}{n}} \Road^\prime\),
since \(|\laneReservation(B)| \neq 2\). 
But, if we let 
\(\Road^\prime = (\laneReservation^\prime ,\laneClaim^\prime, \pos, \spd, \acc)\)  
where \(\laneReservation^\prime\) and \(\laneClaim^\prime\) coincide with
their counterparts in \(\Road\) except for \(\laneReservation^\prime(E) = \{1,2\}\)
and \(\laneClaim^\prime(E) = \emptyset\), then \(\Road \transition{\moveToLane{E}} \Road^\prime\).
\end{exa}

EMLSL restricts the parts of the motorway perceived by each car
to so called \emph{views}. Each view comprises a set of lanes and
a real-valued interval, its length. 

\begin{defi}[View]
\label{def:view}
For a given traffic snapshot \(\Road\) with a set of lanes \(\Lanes\), a 
\emph{view} $V$ is defined as a structure
$V=(\ViewLanes, \Extension, E)$,
where
\begin{itemize}
  \item $\ViewLanes=[l,n]\subseteq \Lanes$ is an interval of lanes that are visible
    in the view,
  \item $\Extension={[r,t]} \subseteq \R$ is the extension that is visible in 
    the view,
  \item $E\in \ID$ is the identifier of the car under consideration, the \emph{owner of the view}.
\end{itemize}
A \emph{subview} of \(V\) is obtained by restricting the lanes and extension we 
observe. For this we use sub- and superscript notation:
$V^{\ViewLanes^{\prime}} =(\ViewLanes^{\prime}, \Extension, E)$
and $V_{\Extension^{\prime}} =(\ViewLanes, \Extension^{\prime}, E)$,
where \(\ViewLanes^{\prime}\) and \(\Extension^{\prime}\) are subintervals of 
\(\ViewLanes\) and \(\Extension\), respectively. 

\end{defi}

While views define the range of the car's sensors, we use a distinct function
to model the capability of these sensors. That is, the perceived length of cars
can be dependent on the car under consideration. As an example, a car may 
calculate its own braking distance, while it can only perceive 
the physical size of all other cars.

\begin{defi}[Sensor Function]
The car dependent \emph{sensor function} 
\(\Omega_E: \ID\times\Roads\rightarrow \R_+\)  given a car identifier and a traffic snapshot provides the length of the corresponding car, as perceived by \(E\). 
\end{defi}

The intention of the sensor function is to parametrize the knowledge available
to the cars and by that to easily allow for the consideration of different scenarios 
\cite{HLOR2011}.

\begin{rem}[Abbreviations]
For a given view $V=(\ViewLanes, \Extension, E)$ and a \tsText{} $\RoadDef$ we use the following abbreviations:
\begin{align*}
  \laneReservation_V: \ID \rightarrow \Powerset{\ViewLanes}\ \text{with}&\ C \mapsto \laneReservation(C) \cap \ViewLanes\\
  \laneClaim_V: \ID \rightarrow \Powerset{\ViewLanes}\ \text{with}&\ C
  \mapsto \laneClaim(C)\cap \ViewLanes\\
  len_V:\ID \rightarrow  \Powerset{\Extension}\ \text {with}&\ C \mapsto
  [pos(C),pos(C)+\oracle{C,\Road}]\cap \Extension
\end{align*}

The functions \(\laneReservation_V\) and \(\laneClaim_V\) are restrictions of their counterparts in \(\Road\) to the sets of lanes considered in this view. 
The function \(len_V\) gives us the part of the  view occupied by a 
car \(C\).\footnote{This presentation differs slightly from the first presentation of MLSL in two ways. First, we do not restrict the set of identifiers anymore to the cars ``visible'' to \(E\). Since the functions for the reservations, claims or length return the empty set for cars outside of \(V\), such cars cannot satisfy the corresponding atomic formulas. The definition of \(\laneReservation_V\) and \(\laneClaim_V\) was altered due to a technical mistake in the previous form. }
\end{rem}

\begin{exa}
To fully formalize Fig.~\ref{fig:example_situation}, we have
to a define a view \(V = (\ViewLanes, \Extension, E)\) corresponding
to the dashed rectangle. 
The set of lanes visible in \(V\) is 
\(\ViewLanes = \{1,2\}\). For the extension, we only have
to choose values such that the relations of the figure are 
preserved, i.e., both \(E\) and \(A\) fit fully into the 
extension, the safety envelope of \(B\) is partially contained
in \(\Extension\), while no part of \(C\) overlaps with it. 
Hence, we first have to define how the safety envelopes
are perceived by \(E\).    
\begin{align*}
    \sensorsFunc{E}{A}{\Road} &= 10  &    \sensorsFunc{E}{B}{\Road} &=11.5  &
    \sensorsFunc{E}{C}{\Road} &= 7  &    \sensorsFunc{E}{E}{\Road} &= 13 
\end{align*}
Now we can choose, e.g.,  \(\Extension = [12, 42]\).
With this view and sensor function, the derived 
functions of \(\Road\) and \(V\) are as follows.
\begin{align*}
 \laneReservation_V(A)& = \{1,2 \}  & \laneReservation_V(B)& = \{1 \}  &
 \laneReservation_V(C)& = \emptyset  & \laneReservation_V(E)& = \{2 \}  \\
 \laneClaim_V(A)& = \emptyset  &  \laneClaim_V(B)& = \emptyset  &
 \laneClaim_V(C)& = \emptyset  &  \laneClaim_V(E)& = \{ 1\}  \\
 \len_V(A) &= [28,38]  & \len_V(B)& = [12,15]  &
 \len_V(C)& = \emptyset  & \len_V(E)& = [14,27]
\end{align*}
Observe how the space occupied by \(B\) is reduced to fit into the view, and that
the reservation of \(C\) is invisible for \(E\), since the view only comprises
both lower lanes.
\end{exa}

In the logic, the view shall be interpreted  relatively to the owner of the view.
If a traffic snapshot \(\Road\) evolves to \(\Road^\prime\) in the time
\(t\), i.e. \(\Road \abstrans{t} \Road^\prime\), the extension \(X\) of a view
\(V = (\ViewLanes, \Extension, E)\) has to be shifted by the difference 
of the positions of \(E\) in \(\Road\) and \(\Road^\prime\). For this purpose,
we introduce the function \(\moveSingle\), which given two snapshots
\(\Road\), \(\Road^\prime\) and a view \(V\) computes the view \(V^\prime\)
corresponding to \(V\) after moving from \(\Road\) to \(\Road^\prime\).

\begin{defi}[Moving a View]
For two traffic snapshots \(\RoadDef\) and 
\(\Road^\prime = (\laneReservation^\prime, \laneClaim^\prime, 
\pos^\prime,\spd^\prime, \acc^\prime)\)  and a view
\( V = (\ViewLanes, [r,s], E)\),  the result of
\emph{moving \(V\) from \(\Road\) to \(\Road^\prime\)} is given
by \(
\move{\Road}{\Road^\prime}{V} = (\ViewLanes, [r + x, s+x], E)  
\),
where \(x = \pos^\prime(E) - \pos(E)\).
\end{defi}

Definition~\ref{def:chop_discrete} formalizes the partitioning of discrete 
intervals. We need this slightly intricate notion to have a clearly defined
chopping operation, even on the empty set of lanes. We want the empty
set to be a valid interval of lanes, so that the smallest intervals of lanes
and horizontal space behave similarly. 
\begin{defi}[Chopping discrete intervals]
\label{def:chop_discrete}
Let \(I\) be a discrete interval, i.e., \(I = [l,n]\) for some
\(l, n \in \Lanes\) or \(I = \emptyset\). Then 
\(\vreach{I^1}{I^2}{I} \) if and only if  \(I^1 \cup I^2 = I\), 
\(I^1 \cap I^2 = \emptyset\), and
both \(I^1\) and \(I^2\) are discrete convex intervals, which implies 
\(\max (I^1) +1 = \min(I^2)\) or  \(I^1 = \emptyset\) or
 \(I^2 = \emptyset\).
\end{defi}

We define the following relations on views to have a consistent description
of vertical and horizontal chopping operations. 
\begin{defi}[Relations of Views]
\label{def:view_relations}
Let \(V_1\), \(V_2\) and \(V\) be views of a snapshot \(\Road\). Then
\( \vreach{V_1}{V_2}{V}\) if and only if \(V = (L,X,E)\), \(\vreach{L_1}{L_2}{L}\), 
\(V_1 = V^{L_1}\) and \(V_2 = V^{L_2}\). 
Furthermore, \(\hreach{V_1}{V_2}{V}\) if and only if \(V = (L,[r,t],E)\) and there
is an \(s \in [r,t]\) such that \(V_1 = V_{[r,s]}\) and \(V_2 = V_{[s,t]}\).  
\end{defi}

To abstract from the borders of real-valued intervals during the definition of the 
semantics, we define the following norm giving the length of such
intervals. This notion coincides with the length measurement of 
DC~\cite{CHR1991}. We also define the cardinality of discrete intervals
to be their length.
\begin{defi}[Measures of intervals]
Let \(I_R= [r,t]\) be a real-valued interval, i.e. \(r,t \in \R\). The
\emph{measure} of \(I_R\) is the norm \(\|I_R\| = t -r \).
For a discrete interval \(I_D\), the measure of \(I_D\)  is simply its
cardinality \(|I_D|\).
\end{defi}

With the definition of measures, 
we can give the reason for the need of Def.~\ref{def:chop_discrete}.
The smallest intervals in horizontal direction  are point-intervals, e.g.
\(I = [r,r]\) for some \(r\in\R\). The measure of \(I\) is \(\|I\| = 0\).
In contrast, if the smallest intervals of lanes were also point-intervals, i.e.,
 sets of the form \(\{n\}\), their measure would be \(|\{n\}|=1\). However, 
with the the empty set as the smallest interval of lanes, the measures behave
similarly for both directions.

We employ three sorts of variables. The set of variables ranging 
over car identifiers is 
denoted by \(\carvariables\), with typical elements \(c\) and \(d\).  
For referring to lengths and quantities of lanes, we use the sorts 
\(\realvariables\) and \(\lanevariables\)
ranging over real numbers and elements of the set of lanes \(\Lanes\), respectively. 
The set of all variables is denoted by \(\variables\). To refer to the car 
owning the current view, we use the special constant
\(\ego\). Furthermore we use the syntax \(\length\) for the length of a
view, i.e., the length of the extension of the view and \(\width\) for
the width, i.e., the number of lanes. For simplicity, we only allow for
addition between correctly sorted terms. However, it is straightforward
to augment the definition with further arithmetic operations.

\clearpage
\begin{defi}[Syntax]
\label{def:syntax}
We use the following definition of \emph{terms}.
  \begin{align*}
    \theta & ::= n \mid r \mid \ego \mid u \mid \length \mid \width \mid \theta_1 + \theta_2,
  \end{align*}
  where \(n \in \Lanes\), \(r\in \R\) and \(u \in \variables\) and \(\theta_i\)
are both of the same sort, and not elements of \(\carvariables\cup\{\ego\}\). 
We denote the set of terms with \(\functerms\).
The syntax of the \emph{extended multi-lane spatial logic EMLSL} is given 
as follows.
  \begin{align*}
    \phi &::= \bot \mid \theta_1 = \theta_2 \mid  \reserved{c} \mid \claimed{c} \mid \phi_1 \implies \phi_2 \mid \forall z \qsep \phi_1 
\mid \phi_1 \chop \phi_2 \mid \begin{array}{c} \phi_2 \\ \phi_1
    \end{array} \mid M \phi
  \end{align*}
where \(M \in \{\boxmodal{\lndres{c}}, \boxmodal{\lndclaim{c}}, \boxmodal{\lndwdclaim{c}},
\boxmodal{\lndwdres{c}}, \boxmodal{\tau} 
\}\), 
 \(c \in \carvariables \cup\{\ego\}\), \(z \in \variables\), and
 \(\theta_1, \theta_2 \in \functerms\) are of the same sort.
We denote the set of all EMLSL formulas by \(\formulae\).
 \end{defi}

\begin{defi}[Valuation and Modification] 
A \emph{valuation} is a function \\
\(\val \colon \variables\cup\{\ego\} \to \ID\cup\R\cup\Lanes \). 
We silently assume valuations and their modifications
to respect the sorts of variables. For a view \(V = (\ViewLanes, \Extension, E)\),
we lift  
\(\val\) to a function \(\val_V\) evaluating terms, where
variables and \(\ego\) are interpreted as  in \(\val\), 
and \(\val_V(\length) = \| \Extension\|\) and
\(\val_V(\width) = | \ViewLanes|\). The function \(+\) is interpreted as addition.
\end{defi}

\begin{defi}[Semantics] \label{def:semantics-MLSL}
In the following, let \(\theta_i\)  be terms of the same sort, 
\(c \in \carvariables\cup\{\ego\}\) 
and \(z \in \variables\).  
The \emph{satisfaction} of formulas 
with respect to a traffic snapshot \(\Road\), 
a view \(V = (\ViewLanes,\Extension,E)\)
and a valuation \(\val\) with \(\val(\ego) = E\)  
is defined inductively as follows:
{ 
\allowdisplaybreaks
\begin{align*}
\Road,V,\val&\not\models \bot & & \phantom{=}\text{ for all } \Road, V, \val\\
\Road,V,\val&\models \theta_1 = \theta_2 & \Leftrightarrow & \phantom{=}\val_V(\theta_1) = \val_V(\theta_2)\\
\Road,V,\val&\models \reserved{c} & \Leftrightarrow & \phantom{=}|\ViewLanes|=1 \text{ and } \|\Extension\| > 0 \text{ and } \\
&&& \phantom{=}\laneReservation_V(\val(c))=\ViewLanes \text{ and } \Extension=len_V(\val(c))\\ 
\Road,V,\val&\models \claimed{c} & \Leftrightarrow & \phantom{=}|\ViewLanes|=1 \text{ and } \|\Extension\| > 0 \text{ and }  \\
&&& \phantom{=}\laneClaim_V(\val(c))=\ViewLanes \text{ and } \Extension=\len_V(\val(c))\\
\Road, V ,\val &\models \phi_1 \implies \phi_2 &\Leftrightarrow & \phantom{=}\Road, V,\val \models \phi_1 \text{ implies } \Road, V,\val \models \phi_2\\
\Road, V, \val &\models \forall z \qsep \phi & \Leftrightarrow&\phantom{=} \forall\, \alpha \in \ID\cup\R\cup\Lanes  \qsep \Road, V, \val\subst{z}{\alpha} \models \phi  \\
\Road,V,\val&\models \phi_1\chop\phi_2&\Leftrightarrow &\phantom{=} 
\exists V_1,V_2 \qsep \hreach{V_1}{V_2}{V} \text{ and }\\
&&&\phantom{=}\Road,V_{1}, \val\models \phi_1\text{ and } \Road,V_{2},\val\models \phi_2\\ 
\Road,V, \val&\models \begin{array}{c}{\phi_2}\\{\phi_1}\end{array}&\Leftrightarrow & \phantom{=} \exists V_1,V_2 \qsep \vreach{V_1}{V_2}{V} \text{ and } \\
&&&\phantom{=}\Road,V_1,\val\models \phi_1  \text{ and } \Road,V_2,\val\models \phi_2\\
\Road, V, \val & \models \boxmodal{\lndres{c}} \phi & \Leftrightarrow & \phantom{=}\forall \Road^\prime \qsep \Road \transition{\moveToLane{\val(c)}} \Road^\prime \text{ implies } \Road^\prime, V, \val \models \phi\\
\Road, V, \val & \models \boxmodal{\lndclaim{c}} \phi & \Leftrightarrow & \phantom{=}\forall \Road^\prime, n \qsep \Road \transition{\claimLane{\val(c)}{n}} \Road^\prime \text{ implies } \Road^\prime, V, \val \models \phi\\
\Road, V, \val & \models \boxmodal{\lndwdclaim{c}} \phi & \Leftrightarrow & \phantom{=}\forall \Road^\prime \qsep \Road \transition{\stopclaimLane{\val(c)}} \Road^\prime \text{ implies } \Road^\prime, V, \val \models \phi\\
\Road, V, \val & \models \boxmodal{\lndwdres{c}} \phi & \Leftrightarrow & \phantom{=}\forall \Road^\prime, n \qsep \Road \transition{\stopmoveToLane{\val(c)}{n}} \Road^\prime \text{ implies } \Road^\prime, V, \val \models \phi\\
\Road, V, \val & \models \boxmodal{\tau} \phi & \Leftrightarrow & \phantom{=}\forall \Road^\prime, t \qsep \Road \abstrans{t} \Road^\prime \text{ implies } \Road^\prime, \move{\Road}{\Road^\prime}{V}, \val \models \phi\\
\end{align*}
}
\end{defi}

Observe that views are only moved whenever time passes between snapshots.
In addition to the standard abbreviations of the remaining Boolean 
operators and the existential quantifier, we use \(\true \equiv \lnot \bot\).
 An important derived 
modality of our previous work \cite{HLOR2011} is 
the \emph{somewhere} modality 
\begin{align*}
\somewhere{\phi} \equiv \true \chop \left(\begin{array}{c} \true \\ \phi \\ \true \end{array}\right) \chop \true .
\end{align*}
 
Further, we use its dual operator \emph{everywhere}.
We abbreviate the modality \emph{somewhere along the extension
of the view} with the operator \(\diamodal{\length}\), similar to 
the \emph{on some subinterval} modality of DC.

\begin{center}
  
\begin{tabular}{ccc}
\(   \everywhere{\phi} \equiv\lnot \somewhere{\lnot\phi}\)\hspace*{2em}&
\( \diamodal{\length} \phi \equiv\true \chop \phi \chop \true\)\hspace*{2em}&
\(\boxmodal{\length} \phi  \equiv\lnot \diamodal{\length} \lnot \phi\)
\end{tabular}
\\
\end{center}

Likewise, abbreviations can be defined to express the modality
\emph{on some lane}. 
Furthermore, we define the diamond modalities
for the transitions as usual, i.e., 
\(
\diamodal{\ast} \phi \equiv  \lnot \boxmodal{\ast} \lnot \phi\),
where \(\ast \in \{\lndres{c}, \lndclaim{c}, \lndwdres{c}, \lndwdclaim{c}, \tau\}\).

In the first definition of MLSL, we included the atom \(\free\) to denote
free space on the road, i.e., space which is  neither occupied by a reservation nor
by a claim. It was not possible to derive this atom from the others, since
we were unable to express the existence of exactly one lane
and a non-zero extension in the view. 
However, in the current presentation,
\(\free\) can be defined within EMLSL. Observe that a view of non-zero extension  
 can be characterized by \(\length > 0 \equiv \lnot (\length = 0)\).
\begin{align*}
  \free & \equiv \length > 0 \land \width = 1 \land \forall c \qsep \boxmodal{\length} (\lnot \claimed{c} \land \lnot \reserved{c})
\end{align*}

Furthermore, we can define  
\(\length < r \equiv \lnot (\length = r \chop \true)\) and use  
the superscript \(\varphi^r\) to abbreviate the schema 
\(\varphi \land \length = r\). For reasons of clarity, we will not always use
this abbreviation and write out the formula instead, to emphasize the restriction.

As an example, the following formula defines the behavior of a safe distance 
controller, i.e., as long as the car starts in a situation with free space in 
front of it,
the formula demands that after an arbitrary time, there is still free space
left. 
\begin{align*}
  \forall  x,y \qsep  \Diamond_\length \parenthesis{\vchop{\width = x}{\vchop{\reserved{\ego}\chop \free}{\width = y}}} \implies \boxmodal{\tau}\parenthesis{\Diamond_\length \parenthesis{\vchop{\width = x}{\vchop{\reserved{\ego}\chop \free}{\width = y}}}}
\end{align*}

We have to relate the lane in both the antecedent and the 
conclusion by the atoms \(\width = x\) and \(\width = y \) respectively. 
If we  simply used \(\somewhere{\reserved{\ego}\chop \free}\), it would be possible
for the reservations  to be on different
lanes, and hence, we would not ensure that free space is in front
 of each of \(\ego\)'s  reservations at every point in time. However,
the formula does not constrain how the situations may change, whenever
reservations or claims are created or withdrawn. 

Observe that it is crucial to combine acceleration and 
time transitions into a single modality~\(\boxmodal{\tau}\). Let \(\ego\) drive
on lane \(m\) with a velocity of \(v\). If we only allowed 
for the passing of time, this formula would require all cars on \(m\)
in front of
\(\ego\) to have a velocity \(v_f \geq v\), while
all cars behind \(\ego\) had to drive with \(v_b \leq v\). 
Hence the evolutions
 allow for more 
complex behavior in the underlying model.

Like for ITL~\cite{Mos1985} or DC~\cite{CHR1991}, 
we call a formula
\emph{flexible} whenever its satisfaction is dependent on the 
 current traffic snapshot and view. Otherwise the formula
is \emph{rigid}. However, since the spatial dimensions of EMLSL are
not directly interrelated, we also distinguish \emph{horizontally rigid}
and \emph{vertically rigid} formulas. The satisfaction of the former
is independent of the extension of views, while for the latter, the
amount of lanes in a view is of no influence.
If a formula is only independent of the current traffic snapshot, 
we call it \emph{dynamically rigid}. 

\begin{defi}[Types of Rigidity]
Let \(\phi\) be a formula of EMLSL. We call \(\phi\) \emph{dynamically rigid},
if it does not contain any spatial atom, i.e., \(\reserved{c}\) or \(\claimed{c}\) 
as a subformula. Furthermore, we call \(\phi\) \emph{horizontally rigid}, if
it is dynamically rigid and in addition 
does not contain \(\length\) as a term. Similarly, \(\phi\) is 
\emph{vertically rigid}, if it is dynamically rigid and does not contain
\(\width\) as a term. If \(\phi\) is both vertically and
horizontally rigid, it is simply \emph{rigid}. 
\end{defi}

\begin{lem}
\label{lem:rigidity}
Let \(\phi\) by dynamically rigid and \(\phi_H\) (\(\phi_V\)) be horizontally 
(vertically) rigid.
Then for all 
traffic snapshots \(\Road\), \(\Road^\prime\), views \(V\), \(V_1\), \(V_2\) 
and valuations \(\val\),
\begin{enumerate}
\item \(\Road, V,\val \models \phi\) iff \( \Road^\prime, V,\val \models \phi\)
\item Let \(\hreach{V_1}{V_2}{V}\). Then \(\Road, V, \val \models \phi_H\) iff
\(\Road, V_i, \val \models \phi_H\) (for \(i \in \{1,2\}\)).
\item Let \(\vreach{V_1}{V_2}{V}\). Then \(\Road, V, \val \models \phi_V\) iff
\(\Road, V_i, \val \models \phi_V\) (for \(i \in \{1,2\}\)).
\end{enumerate}
\end{lem}

\proof
By induction on the structure  of EMLSL formulas.
\qed


\section{Undecidability of pure MLSL}
\label{sec:undec}

In this section we give an undecidability result for the spatial
fragment of EMLSL, i.e., we do not need the modalities for the 
discrete state changes of the model or the evolutions.
We will call
this fragment \emph{spatial MLSL}, subsequently. 
We reduce the halting problem of two-counter machines,
which is known to be undecidable \cite{Minsky1967}, to satisfaction of 
spatial MLSL formulas.

Intuitively, a two counter machine executes a branching program 
which manipulates a (control) state and  increments and decrements 
two different counters \(c_1\) and
\(c_2\).
Formally, two counter machines consist of a set of states 
\(Q = \{q_0, \dots, q_m\}\), distinguished initial and final states 
\(q_0, q_{\mathit{fin}} \in Q\) and a set of instructions \(I\) of
the form shown in Tab.~\ref{tab:two-counter-instructions} (the instructions
for the  counter \(c_2\) are analogous). The
instructions mutate configurations of the form \(s = (q_i, c_1, c_2)\),
where \(q_i \in Q\) and \(c_1, c_2 \in \N\) 
into new configurations:

\begin{table}[ht]
  \centering
  \caption{Instructions for counter \(c_1\) of a two-counter machine}
  \begin{tabular}{|l|l|l|}
\hline
\(s\) & Instruction & \(s^\prime\)\\
   \hline
   \((q, c_1, c_2)\) & \(q \transition{c_1^+} q_j\) & \((q_j, c_1+1, c_2)\) \\
 \((q, 0,c_2)\) & \(q \transition{c_1^-} q_j,q_n\) & \((q_j,0,c_2)\)\\
 \((q, c+1,c_2)\) & \(q \transition{c_1^-} q_j,q_n\) & \((q_n,c,c_2)\)\\
\hline
  \end{tabular}
  \label{tab:two-counter-instructions}
\end{table}

An \emph{run from the initial configuration} of a two-counter 
machine \((Q,q_0,q_{\mathit{fin}}, I)\) is a sequence of configurations
\((q_0,0,0) \transition{i_0} \dots \transition{i_p} (q_{p+1}, c_{p+1}, c^\prime_{p+1})\),
 where each \(i_j\) is an instance of an instruction within \(I\). If 
\(q_{p+1} = q_{\mathit{fin}}\), the run is \emph{halting}.

We  follow the approach of 
Zhou et al. \cite{CHS1993} for DC. They encode the configurations 
in recurring patterns of length \(4k\), where the first
part constitutes the current state, followed by the contents
of the first counter. The third part is filled with a marker
to distinguish the counters, and is finally followed by 
the contents of the second counter. Each of these parts
is exactly of length \(k\).

Zhou et al. could use distinct observables 
for the state of the machine,  counters and separating
delimiters, since DC allows for the definition
of arbitrary many observable variables. We have to modify this 
encoding since within spatial MLSL we are restricted to two 
 predicates for reservations and claims, and the derived predicate for 
free space, respectively. Furthermore, due to the constraints
on EMLSL models in Def.~\ref{def:snapshots}, we cannot use 
multiple occurrences of reservations of a unique
car to stand, e.g., for the  values of one counter. 
Hence we have to existentially quantify all mentions of reservations
and claims. We will never reach an upper limit of existing cars,
since we assume \(\ID\) to be countably infinite.

The current state of the machine \(q_i\) is encoded by the number
of lanes below the current configuration,
the states of the counters is described by a sequence of reservations,
separated by a single claim. To safely refer to the start of a 
configuration, we also use an additional marker consisting of 
a claim, an adjacent reservation and again a claim. 
Each part 
of the configurations is assumed to have length \(k\). Free space
separates the reservations within one counter from each other and
from the delimiters. 
Intuitively, a configuration is encoded as follows:
\begin{center}
\includegraphics[scale=.85]{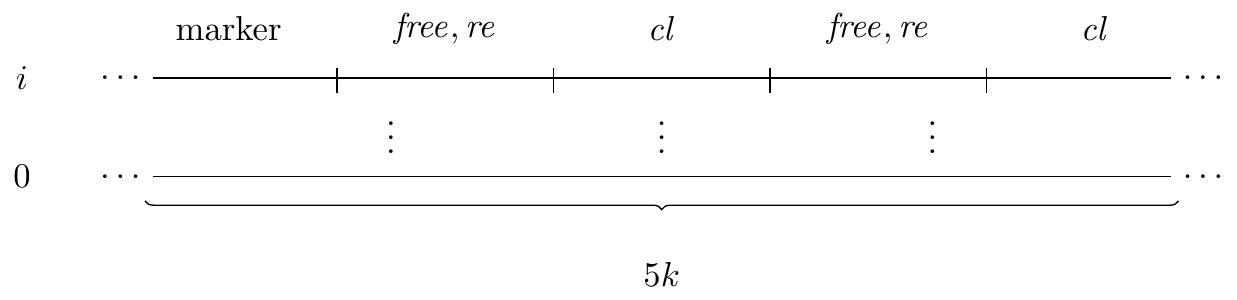}
\end{center}
To enhance the readability of our encoding, we use the 
abbreviation 
\begin{align*}  
\marker \equiv \exists c \qsep \claimed{c}\chop \exists c \qsep \reserved{c} \chop
\exists c \qsep \claimed{c}
\end{align*}
to denote the start of a configuration.

Like Zhou et al., we ensure that reservations and claims are
mutually exclusive. We do not have to consider \(\free\), since
it is already defined as the absence of both reservations and claims.
Observe that we use the square brackets to denote the \emph{everywhere} modality 
(cf. Section~\ref{sec:mlsl}).
\begin{align*}
  \mutex &= \forall c,d \qsep \everywhere{\claimed{c} \implies \lnot \reserved{d}) \land
(\reserved{c} \implies \lnot \claimed{d}}.
\end{align*}

The initial marking \((q_0, 0,0)\) is then defined by the following formula.
\begin{align*}
  \init &= \parenthesis{\vchop{\vchop{\everywhere{\lnot \exists c \qsep \claimed{c}}}{\marker^k \chop \free^k \chop (\exists c \qsep \claimed{c})^k \chop \free^k \chop (\exists c \qsep \claimed{c})^ k}}{\width = 0}} \chop \true
\end{align*}

We have to ensure that the configurations occur periodically after every \(5k\) spatial
units. Therefore, we use the following schema \(\mathit{Per}(\mathcal{D})\). Observe that 
we only require that the lanes surrounding the formula \(\mathcal{D}\) do not contain
claims. This ensures on the one hand that no configuration lies in parallel with
the formula \(\mathcal{D}\), since well-defined configurations have to include claims.
On the other hand, it allows for satisfiability of the formula, 
since we do not forbid the occurrence
of reservations, which are needed for the claims within the configurations.
\begin{align*}
 \Per{\mathcal{D}} &= \everywhere{\parenthesis{\vchop{\vchop{\everywhere{\lnot\exists c \qsep \claimed{c}}}{\mathcal{D}}}{\everywhere{\lnot\exists c \qsep \claimed{c}}}
\chop \length = 5k} 
\implies 
\parenthesis{ \length = 5k \chop
  \vchop{\vchop{\everywhere{\lnot\exists c \qsep \claimed{c}}}{\mathcal{D}}}{\everywhere{\lnot\exists c \qsep \claimed{c}}}}
}
\end{align*}

Note that we did not constrain on which lane the periodic behavior occurs. This 
will be defined by the encoding of the operations. 

Now we may define the periodicity of the delimiters and  the counters. Here we 
also have to slightly deviate from Zhou et al.: we are not able to express the
statement ``almost everywhere \free{} or \reserved{c} holds,'' directly. We
have to encode it by ensuring that on every subinterval with a length greater
than zero, we can find another subinterval which satisfies \(\free\) or 
\(\reserved{c}\). This expresses in particular, that no claim may occur, due
to the mutual exclusion property.
\begin{align*}
  \periodic &= \Per{(\Box_\length(\length >0 \implies \true \chop (\free \lor \exists c \qsep \reserved{c}) \chop \true ) \land \width = 1)^k }\\
&\land \Per{(\exists c \qsep \claimed{c})^k}  \land \Per{\marker^k}
\end{align*}
  
We turn to the encoding of the operation \(q_i \xrightarrow{c_1^+} q_j\), i.e.,
the machine goes from \(q_i\) to \(q_j\) and increments the first counter by one.
Similar to Zhou et al., we use encodings of  the form 
\(\lnot (\mathcal{D}_1 \chop \lnot \mathcal{D}_2)\), meaning ``whenever the 
beginning of the view satisfies \(\mathcal{D}_1\), the next part satisfies 
\(\mathcal{D}_2\).''

The formula \(F_1\) copies the reservations of counter one of state \(q_i\)
to the corresponding places in counter one in state \(q_j\).
\makeatletter
\newcommand{\vast}{\bBigg@{3}}
\newcommand{\Vast}{\bBigg@{4}}
\makeatother
\begin{align*}
  F_1 &= \lnot \vast(\parenthesis{\vchop{\vchop{\true}{
\marker^k \chop \length < k \chop\exists c \qsep \reserved{c} \chop
((\exists c \qsep \reserved{c} \chop \true) \land \length = 5k)
}}{\width = i}} \chop\\
&\phantom{=}\quad \lnot \parenthesis{\vchop{\vchop{\true}{\length = 0 \lor (\exists c \qsep \reserved{c} \chop \true)}}{\width = j}} \vast)
\end{align*}

We use a similar formula \(F_{free}\) to copy the free space before the reservations.

The formulas \(F_2\) and \(F_3\) handle the addition of another reservation to 
the counter. We have to distinguish between an empty counter and one
already containing reservations.
\begin{align*}
  F_2 &= \parenthesis{\vchop{\vchop{\true}{\marker^k \chop \free^k \chop \length = 5k}}{\width = i}} \implies \parenthesis{\vchop{\vchop{\true}{\true \chop (\free \chop \exists c \qsep \reserved{c} \chop \free)^k}}{\width = j}}
\end{align*}

\begin{align*}
  F_3 &= \parenthesis{\vchop{\vchop{\true}{\marker^k \chop \length < k \chop \exists c \qsep \reserved{c} \chop ((\free\chop \exists c \qsep \claimed{c} \chop \true) \land \length = 6k)}}{\width = i}} \implies \\
&\phantom{=}\quad \parenthesis{\vchop{\vchop{\true}{\true \chop (\free \chop \exists c \qsep \reserved{c} \chop \free \chop \exists c \qsep \claimed{c})^k}}{\width = j}} 
\end{align*}

In addition, we need formulas which copy of contents of the
second counter to the new configuration, similar to \(F_1\).

Let \(I_C\) be the set of the machine's instructions and  \(F(i)\) 
be the conjunction of the formulas encoding operation \(i\)
and \(q_{\mathit{fin}}\) its final state.
Then 
\begin{align*}
  \halt{C} &= \init \land \periodic \land \mutex \land 
\bigwedge_{i \in I_C}\Box_\length F(i) \land \Diamond_\length \parenthesis{\vchop{\vchop{\true}{\exists c \qsep \claimed{c}}}{\width = \mathit{fin}}}.
\end{align*}

If and only if \(\halt{C}\) is satisfiable, the machine contains a halting run.
This holds since only configurations may contain claims (as defined in the
formalization of periodicity), and whenever the machine reaches its final
state, it halts. Hence the halting problem of two counter machines with empty 
initial configuration reduces to satisfiability of spatial MLSL formulas.
\begin{prop}
\label{prop:reduct}
  Let \(C\) be a two counter machine. Then \(C\) has a halting run
if and only if \(\halt{C}\) is satisfiable. 
\end{prop}
\proof\hfill
  
\emph{``if''}. 

Let \(\Road, V, \val \models \halt{C}\), where 
\(V = (\ViewLanes, \Extension, E)\). Observe that all 
variables occurring in \(\halt{C}\) are existentially quantified, and
hence we may ignore the values of \(\val\). We divide \(\Extension\) into
parts of length \(5k\), i.e., we have \(|\Extension| = s \cdot 5k + r\),
where \(0 \leq r < 5k\), which means
\begin{align*}
\Extension = [a,b] = \bigcup_{d = 1}^s [a+(d-1)\cdot5k, a+d\cdot 5k] \cup [a+s\cdot 5k, b].
\end{align*}
We denote \(\bigcup_{d = 1}^e [a+(d-1)\cdot5k, a+d\cdot 5k]\) by \(X_e\).
Let \(X^\prime = [a+(d^\prime-1)\cdot 5k, a+d^\prime\cdot 5k]\) and \(X^{\prime\prime} = [a+d^\prime\cdot 5k, a+(d^\prime+1)\cdot 5k]\) for some \(0 <d^\prime < s\).
Now assume that at \(X^\prime\),  lane \(m\) contains 
 a configuration, i.e., 
\begin{align*}
\Road, V_{X^\prime}^{\{m\}} &\models \marker^{k} \chop  (\Box_\length(\length >0 \implies \true \chop (\free \lor \exists c \qsep \reserved{c}) \chop \true ) \land \width = 1)^k \\
&\chop \exists c \qsep \claimed{c}^k \chop (\Box_\length(\length >0 \implies \true \chop (\free \lor \exists c \qsep \reserved{c}) \chop \true ) \land \width = 1)^k \\
&\chop \exists c \qsep \claimed{c}^k
\end{align*}
By interpreting \periodic{} on \(\Road,V_{X^\prime\cup X^{\prime\prime}}\) we get that there
is a lane \(m^\prime\) such that 
\begin{align*}
\Road, V_{X^{\prime\prime}}^{\{m^\prime\}} &\models \marker^{k} \chop  (\Box_\length(\length >0 \implies \true \chop (\free \lor \exists c \qsep \reserved{c}) \chop \true ) \land \width = 1)^k \\
&\chop \exists c \qsep \claimed{c}^k \chop (\Box_\length(\length >0 \implies \true \chop (\free \lor \exists c \qsep \reserved{c}) \chop \true ) \land \width = 1)^k \\
&\chop \exists c \qsep \claimed{c}^k
\end{align*}

Furthermore, \(\periodic\) prevents  that there exists a lane different
from \(m^\prime\) containing such a situation, since for it to hold, all other
lanes are forbidden to contain claims at \(X^{\prime\prime}\). Hence we have
exactly one configuration on all parts \([a+(d-1)\cdot 5k, a+d\cdot 5k ]\).
 
We can extract a 
run for \(C\) from \(\Road, V\) from \(\halt{C}\) by 
induction on \(d\) as follows. 

Let \(d = 1\). Then \(\init\) ensures that on lane \(0\), there is a configuration
with no reservations between \(\marker\) and the first claim and between
the first and the second claim. Hence, we have a run starting at and ending with
 \((q_0,0,0)\).

As the induction hypothesis, we assume that for \(1 \leq d < s\), we can extract a run  
\(R = (q_0,0,0) \transition{}^\ast(q_i,c_1,c_2)\) from \(\Road, V_{X_d}\). For \(d+1\), we know by the 
arguments above, that there exists exactly one configuration on 
\([a+d\cdot 5k, a+(d+1) \cdot 5k]\). Since \(C\) is deterministic, 
for the configuration on lane \(i\), there is at most one set of formulas
applicable.
We only show the case for instruction  incrementing  counter one.

 Let \(F_1, F_2, F_3, F_\free\) be the applicable formulas, 
which we will interpret on \(X_{d+1} \setminus X_{d-1}\), i.e. the
interval \(X_{+} = [a+(d-1)\cdot 5k, a+(d+1)\cdot 5k]\). This
interval is exactly \(10k\) long and starts with \(\marker^k\) on lane
\(i\). Then \(F_1\) states that for each reservation in the representation
of the first counter, i.e., where \(\length < k \chop \exists c\qsep \reserved{c}\)
holds, we
find a reservation on lane \(j\) exactly \(5k\) space units onwards. The 
outermost negation ensures that each possible chop point is considered, 
in particular the chop points arbitrarly close to the end points of the 
reservations. \(F_\free\) ensures in a similar way, that for each free space in
front of a reservation in this representation, we have free space exactly \(5k\)
space units onwards on lane \(j\). Hence, all reservations and the free space in
between is present on lane \(j\). 

Now we consider two cases. When there
is no reservation between the marker and the first single claim, then
\(F_2\) replaces this free space by a reservation enclosed by free space,
i.e., the end configuration of the run was  \((q_i,  0, c_2)\) and the
resulting configuration is \((q_j,1,c_2)\). The second counter was 
copied like the first. 

If there was a reservation before the last free space, then \(F_3\) 
replaces this last free space similarly by a reservation enclosed by free
space on lane \(j\), i.e., the configuration \((q_i,c_1,c_2)\) is changed 
to \((q_j,c_1+1,c_2)\). Hence, in both cases we defined the increment of 
counter 1 together with a state change from \(q_i\) to \(q_j\), which is
by construction an instruction of \(C\), hence \(R \transition{} (q_j, c_1,c_2)\)
is a valid run of \(C\).
The other cases are analogous. 

Now if we did extract a run from the 
satisfying model of \(\halt{C}\), we have two possibilities. 
First, if \(r = 0\), then the configuration at step \(s\) is the last 
of \(R\). Then the last conjunct of \(\halt{C}\) ensures, that
a final state was reached, hence \(R\) is a halting run.

Otherwise, if \(r > 0\), then similarly it is ensured that on this last
part of \(V\),  the lane corresponding to the final state has been reached.
Since also the last change has to be initiated by a formula as before,
there is an instruction to complete \(R\) to a halting run. 
  
\emph{``only if''}.

Let \(R = (q_0,0,0) \transition{}^\ast (q_{fin},c_1,c_2)\) be a halting
run of \(C\) with \(d+1\) configurations, i.e. \(q_d = q_{fin}\). 
We create a model \(\Road, V\) with \(V = (\ViewLanes, \Extension, E)\)
with \(|\Extension| = (d+1) \cdot 5k\) and \(|\ViewLanes| = |Q|+1\) as follows. 
For
a configuration \((q_i, c_1,c_2)\) at step \(d^\prime\), we define 
three cars 
\(C_{d^\prime,0},C_{d^\prime,1},C_{d^\prime,2}\) with
\begin{align*}
  \pos(C_{d^\prime,e}) &= d^\prime \cdot 5k + e \cdot k/3 & \text{ for } e \in \{0,1,2\}\\
  \laneReservation(C_{d^\prime,0}) = \laneReservation(C_{d^\prime,2}) & = \{i+1\}\\
  \laneReservation(C_{d^\prime,1}) & = \{i\}\\
  \laneClaim(C_{d^\prime,0}) = \laneClaim(C_{d^\prime,2}) & = \{i\}\\
  \Omega_E(C_{d^\prime,e},\Road) &= k/3 & \text{ for } e \in \{0,1,2\} 
\end{align*}

These cars satisfy \(\marker^k\). For the claims marking the end of counter
1 and 2 respectively, we define \(C_{d^\prime,4}\) and \(C_{d^\prime,6}\) as follows.
\begin{align*}
  \pos(C_{d^\prime,4}) &= d^\prime \cdot 5k + 2k  \\
  \pos(C_{d^\prime,6}) &= d^\prime \cdot 5k + 4k  \\
  \laneReservation(C_{d^\prime,4}) = \laneReservation(C_{d^\prime,6}) & = \{i+1\}\\
  \laneClaim(C_{d^\prime,4}) = \laneClaim(C_{d^\prime,6}) & = \{i\}\\
  \Omega_E(C_{d^\prime,4},\Road) &= \Omega_E(C_{d^\prime,6}) = k   
\end{align*}

For the definition of the first counter, we need the maximum value \(\mathit{max}\) of both
counters on the whole run. Then we define a sequence of cars \(C_{d^\prime, 3,x}\),
where \(1 \leq x \leq c_1\) if \(c_1 > 0\). For each such car we set

\begin{align*}
  \pos(C_{d^\prime, 3,x}) &= d^\prime \cdot 5k + 3k + \left((2x +1) \cdot \frac{k}{1+ 2\cdot \mathit{max}}\right)\\
\laneReservation(C_{d^\prime, 3,x}) & = \{i\}\\
\laneClaim(C_{d^\prime, 3,x}) & = \emptyset\\
\Omega_E(C_{d^\prime, 3,x}) &= \frac{k}{1+ 2\cdot \mathit{max}}
\end{align*}
Otherwise, no such sequence is added. 

For the second counter, we define a similar sequence \(C_{d^\prime, 5,x}\) with
\( 1 \leq x \leq c_2\) if \(c_2 > 0\). 

If we create such sets of cars for each configuration, the formula \(\halt{C}\) is
satisfied, if the run is halting. \qed

The main theorem of this section is a corollary of Prop.~\ref{prop:reduct}.
\begin{thm}
  The satisfiability problem of spatial MLSL is undecidable.
\end{thm}

Even though we used the full power of spatial MLSL in the proof, i.e.,
we used both \(\length\) and \(\width\), the proof would be possible without
using the latter. For that, we would not be able to encode the state of the
configuration in the lanes, but by a similar way to the markers in the formulas. 
For example, the formula 
\((\exists c \qsep \claimed{c} \chop \exists c \qsep \reserved{c} \chop
\exists c \qsep \claimed{c})^k\) would denote the state \(q_0\), and with
another iteration of \(\reserved{c}\), it would denote \(q_1\) and so on. If
we remove the references to more than one lane in each of the formulas above,
the reservations and claims would already imply that only one lane exists, 
and hence, the use of \(\width\) within the abbreviation \free{} could be
omitted. This shows that spatial MLSL is already undecidable 
even if we only use \(\length\).


\section{Labelled Natural Deduction for EMLSL}
\label{sec:lnd}

Despite the negative decidability result of the previous section,  we define
 a system of labelled natural deduction \cite{Gab1996,BMV1998,Vig2000}
for the full logic EMLSL. 
That is, the rules of the deduction system do not operate on formulas \(\phi\), but
on \emph{labelled formulas} \(\lwff{w}{\phi}\), where \(w\) is a term of a 
\emph{labelling algebra} and \(\phi\) is a formula of EMLSL. 
They may connect the derivations of formulas and relations between the 
terms \(w\)
to 
allow for a tighter relationship between both. 
The labelling algebra is more involved than for standard modal logics,
since EMLSL is in essence a multi-dimensional logic, where the modalities
are not interdefinable. Obviously, the spatial modalities
can not be defined by the dynamic modalities and vice versa. Furthermore,
neither can the dynamic modalities  be defined by each other in general. 
Consider, e.g., the modalities \(\boxmodal{\lndres{c}}\) and 
\(\boxmodal{\lndclaim{c}}\). Both of these modalities rely on different 
transitions between the models, which are only indirectly related.

The labels of the algebra consist of tuples \(\Road, V\), where similar
to the semantics, \(\Road\) is the name of a traffic snapshot and \(V\)
a view. The algebra is then twofold. The relations of the form 
\(\hreach{V_1}{V_2}{V}\) and \(\vreach{V_1}{V_2}{V}\) define
ternary reachability relations between views for the spatial modalities.
Relations between snapshots and views, e.g., \(\Road, V 
\transition{\lndres{C}} \Road^\prime, V\)
describe the behavior of transitions. 
The relations within the labelling algebra for traffic snapshots directly
correspond to the dynamic modalities. For example, we have 
\(\Road, V \transition{\lndclaim{C}} \Road^\prime, V\), whenever there exists an 
\(n \in \N\) such that \(\Road \transition{\claimLane{C}{n}} \Road^\prime\).

We do not give a deduction system
for the transitions between snapshots, since the conditions needed to hold between
them are of a very complex nature, i.e., they are definable only with the 
power of full first-order logic with functions, identity and arithmetic. Hence
we would not achieve a system with a nice distinction between the relational
deductions and the deductions of labelled formulas \cite{BMV1998,Vig2000}.
Furthermore, the possiblity of a transition may be dependent
on properties of cars at any place within the traffic snapshot. This means that we would
have to specify \emph{global} dependencies, while all logical operations we have at
hand are only able to denote \emph{local} properties, i.e., properties
of cars visible in the current view.
Instead we simply assume the existence
of the relations between snapshots whenever needed. That is, we
will often have, e.g., the existence of a transition in our set of assumptions.
This is sensible, since we often want to reason about
  the outcome of a specific transition (see, e.g., Lemma~\ref{lem:reservation}). However, we give simple rules  defining that chopping of a view
into two subviews is always possible.

\begin{defi}[Labelled Formulas and Relational Formulas]
Let \(\Road\) be a name for a traffic snapshot, \(V\) a name 
for a view and \(\phi\) a formula according to Definition~\ref{def:syntax}.
Then \(\lwff{\Road,V}{\phi}\) is a \emph{labelled formula} of EMLSL.
Furthermore, we use two types of \emph{relational formulas}.
On the one hand, we use 
\(\Road, V \transition{\alpha} \Road^\prime, V^\prime\) to denote the existence
of a transition with the label \(\alpha\). On the other hand, the formulas
\(\hreach{V_1}{V_2}{V}\) and
\(\vreach{V_1}{V_2}{V}\) describe that the view \(V\) can be horizontally 
(vertically, resp.) chopped into the views \(V_1\) and \(V_2\).
\end{defi}

To have a meaningful soundness result of the calculus, 
we relate the semantics of labelled formulas with the semantics of normal formulas. 
Observe that we do not define a completely independent notion of models, 
but only use a valuation for this purpose. This is due to the semantic
information which is still comprised within the views and traffic snapshots.
\begin{defi}[Satisfaction of Labelled Formulas]
\label{def:lwff_sat}
We say that a valuation \(\val \)   
\emph{satisfies} a labelled formula
\(\lwff{\Road,V}{\phi}\), written \(\val \models \lwff{\Road, V}{\phi}\) if
and only if \(\Road, V, \val \models \phi\).  Furthermore, 
{
\allowdisplaybreaks
\begin{align*}
\val& \models \Road_1, V \transition{\lndres{c}} \Road_2, V& \metaequiv& \phantom{=}
 \Road_1 \transition{\moveToLane{\val(c)}} \Road_2 ,\\
\val &\models \Road_1, V \transition{\lndwdres{c}} \Road_2, V&\metaequiv &\phantom{=}
\exists n \qsep  \Road_1 \transition{\stopmoveToLane{\val(c)}{n}} \Road_2 ,\\
\val &\models \Road_1, V \transition{\lndclaim{c}} \Road_2, V& \metaequiv& \phantom{=}
\exists n \qsep  \Road_1 \transition{\claimLane{\val(c)}{n}} \Road_2\\
\val &\models \Road_1, V \transition{\lndwdclaim{c}} \Road_2, V&\metaequiv & \phantom{=} 
\Road_1 \transition{\stopclaimLane{\val(c)}} \Road_2\\
\val &\models \Road_1, V_1 \transition{\tau} \Road_2, V_2&\metaequiv& \phantom{=}
\exists t \qsep  \Road_1 \abstrans{t} \Road_2 \text{ and } V_2 = \move{\Road_1}{\Road_2}{V_1}\\
\end{align*}
}
The relational formulas \(\hreach{V_1}{V_2}{V}\) and \(\vreach{V_1}{V_2}{V}\) 
are independent of the valuation at hand, and hence are satisfied whenever 
\(V_1\) and \(V_2\) combined according to Definition~\ref{def:view_relations} 
result in \(V\).

We lift the satisfaction relation also to sets of labelled formulas and 
relational formulas.
Let \(\val\) be a valuation, \(\Gamma\) a set of labelled formulas and
\(\Delta\) a set of relational formulas. Then
\begin{align*}
  \val& \models \Delta& \metaequiv& \phantom{=} 
\forall \rho \in \Delta \qsep \val \models \rho\\
  \val& \models \Gamma& \metaequiv& \phantom{=} 
\forall\; (\lwff{\Road,V}{\phi}) \in \Gamma \qsep \val \models \lwff{\Road,V}{\phi}\\
  \val& \models (\Gamma, \Delta)& \metaequiv& \phantom{=} 
\val \models \Gamma \text{ and } \val \models \Delta\\
\Gamma, \Delta &\models \lwff{\Road, V}{\phi} & \metaequiv& \phantom{=} 
  \val \models (\Gamma, \Delta) \text{ implies } \val \models \lwff{\Road,V}{\phi} \\
&&& \phantom{=} \text{for all valuations } \val
\end{align*}
\end{defi}

\begin{defi}[Derivation]
A \emph{derivation} of a labelled formula \(\lwff{\Road, V}{\phi}\) from
a set of labelled formulas \(\Gamma\) and a set of relational
formulas \(\Delta\) is a tree, where the root is \(\lwff{\Road, V}{\phi}\),
each leaf is an element of \(\Gamma\) or \(\Delta\) and each node
within the tree is a result of an application of one of the rules
defined subsequently. We denote the existence of such a derivation
by \(\Gamma, \Delta \proofs\lwff{\Road, V}{\phi}\).  
\end{defi}

Following Rasmussen \cite{Ras2001}, we define predicates for chop-freeness
of formulas and rigidity of terms and formulas. To increase the 
deducible theorems, we differentiate between \emph{vertical} and \emph{horizontal}
chop-freeness and rigidity. These properties are especially important 
for the correct instantiation of terms, i.e., for the elimination
of universal quantifiers.

\begin{exa}
  Consider the formula
  \begin{align*}
    \forall x \qsep \left(\vchop{\length = x}{\length = x}  \implies \length = x\right),
  \end{align*}
which is a theorem of MLSL, since the length of a view is not changed by chopping
vertically.
If we use classical universal quantifier instantiation and  substitute the  
vertically flexible term \(\width\) for \(x\), then we would get
  \begin{align}
    \vchop{\length = \width}{\length = \width}  \implies \length = \width. \label{ex:1}
  \end{align}
Now let \(V\) be a view satisfying the antecedent of (\ref{ex:1}). 
Then \(V\) can be vertically chopped
such that its length equals its width on both subviews. Now let \(\length = c\).
Then also \(\width = c\) for both subviews. Since \(V\) consists of both these
subviews, \(V\) satisfies \(\width = 2c\). But the conclusion of (\ref{ex:1}) states
that \(V\) should satisfy \(\width = \length = c\). However, we could of course 
substitute
\(x\) by the vertically rigid term \(\length\). 
\end{exa}

We denote vertical (horizontal) chop-freeness by the predicate \(\vcf\) 
(\(\hcf\)) and vertical (horizontal) rigidity by \(\vri\) (\(\hri\)). The rules
for the definition of all four predicates are straightforward, since both
rigidity and chop-freeness are syntactic properties. All atomic formulas
are vertically and horizontally chop-free. For \(\oslash\) being a Boolean operator or
the horizontal chop \(\chop\) , the following
rules give vertical chop-freeness.

\begin{center}
\scalebox{\factor}{
\begin{tabular}{ccc}
\def\defaultHypSeparation{\hskip .1in}
\AXC{\(\vcf(\phi)\)}
\AXC{\(\vcf(\psi)\)}
\RL{\(\vcf \oslash\In\)}
\BIC{\(\vcf(\phi \oslash \psi)\)}
\DisplayProof\hspace{1em}
&
\AXC{\(\vcf(\phi \oslash \psi)\)}
\RL{\(\vcf \oslash\El\)}
\UIC{\(\vcf(\phi)\)}
\DisplayProof\hspace{1em}
&
\AXC{\(\vcf(\phi \oslash \psi)\)}
\RL{\(\vcf \oslash\El\)}
\UIC{\(\vcf(\psi)\)}
\DisplayProof
\\
\end{tabular}}
\end{center}

The  rules for  quantifiers and the 
horizontal rules are defined similarly.

For terms, \(\length\) is vertically
rigid and \(\width\) is horizontally rigid. 
The spatial atoms \(\reserved{c}\) and \(\claimed{c}\) 
are neither horizontally nor vertically rigid, since they require
the view to possess an extension greater than zero and exactly one lane.
Equality is both vertically and horizontally rigid, as long as both
compared terms are rigid. We show some exemplary rules, where
\(\otimes\) is an arbitrary binary operator. 

\begin{center}
\scalebox{\factor}{
\begin{tabular}{ccc}
\AXC{\(\hri(\phi)\)}
\AXC{\(\hri(\psi)\)}
\RL{\(\hri \otimes \In\)}
\BIC{\(\hri(\phi \otimes \psi)\)}
\DisplayProof\hspace{1em}
&
\AXC{\(\hri(\phi \otimes \psi)\)}
\RL{\(\hri \otimes \El\)}
\UIC{\(\hri(\phi)\)}
\DisplayProof\hspace{1em}
&
\AXC{\(\hri(\phi \otimes \psi)\)}
\RL{\(\hri \otimes \El\)}
\UIC{\(\hri(\psi)\)}
\DisplayProof
\\
\end{tabular}}
\end{center}

We have only two simple rules for the relations between views.
First, we state that each view \(V\) is decomposable into two
subviews. This is true, since we allow for the empty view, i.e.,
the view without lanes or with a point-like extension.
 We use \(\vexists\)
to denote  existential quantification over views. 
To use the relations between views, we have to be able to instantiate 
views, i.e., we have to introduce a rule for 
\emph{elimination of existential quantifiers over views}. 
As a side condition for this elimination rule, we require that 
\(\lwff{\Road, V_3}{\phi}\) is not dependent on any assumption including
\(V_1\) or \(V_2\) as a label, except for \(\vreach{V_1}{V_2}{V}\). 
The rule itself is a straightforward adaptation of the classical
rule. Again, we only show the case for the vertical relations.

\begin{center}
\scalebox{\factor}{
\begin{tabular}{cc}
\AXC{}
\RL{VDec}
\UIC{\(\vexists V^\prime, V^{\prime\prime} (\vreach{V^\prime}{V^{\prime\prime}}{V})\)}
\DisplayProof 
&
\AXC{\(\vexists V^\prime, V^{\prime\prime} (\vreach{V^\prime}{V^{\prime\prime}}{V})\)}
\AXC{\([\vreach{V_1}{V_2}{V}]\)}
\noLine
\UIC{\vdots}
\noLine
\UIC{\(\lwff{\Road, V_3}{\phi}\)}
\RL{\(\vexists\El\)}
\BIC{\(\lwff{\Road, V_3}{\phi}\)}
\DisplayProof
\end{tabular}
}
\end{center}

The intuition of rigidity 
is formalized in the following 
rules.  
Whenever a formula is horizontally 
rigid,  the formula holds on all views horizontally
reachable from the current view. Observe that the traffic snapshot may
change arbitrarily, since horizontally rigid formulas are also 
dynamically rigid. The rules for vertically rigidity are similar.

\begin{center}
\scalebox{\factor}{
\begin{tabular}{cc}
\AXC{\(\lwff{\Road, V }{\phi}\)}
\AXC{\(\hri(\phi)\)}
\AXC{\(\hreach{V_1}{V_2}{V}\)}
\RL{\(R_H\)}
\TIC{\(\lwff{\Road^\prime, V_1}{\phi}\)}  
\DisplayProof&
\AXC{\(\lwff{\Road, V}{ \phi}\)}
\AXC{\(\hri(\phi)\)}
\AXC{\(\hreach{V_1}{V_2}{V}\)}
\RL{\(R_H\)}
\TIC{\(\lwff{\Road^\prime,V_2}{ \phi}\)}
\DisplayProof  \\
&\\
\AXC{\(\lwff{\Road, V_1 }{\phi}\)}
\AXC{\(\hri(\phi)\)}
\AXC{\(\hreach{V_1}{V_2}{V}\)}
\RL{\(R_H\)}
\TIC{\(\lwff{\Road^\prime, V}{\phi}\)}  
\DisplayProof&
\AXC{\(\lwff{\Road, V_2}{ \phi}\)}
\AXC{\(\hri(\phi)\)}
\AXC{\(\hreach{V_1}{V_2}{V}\)}
\RL{\(R_H\)}
\TIC{\(\lwff{\Road^\prime,V}{ \phi}\)}
\DisplayProof  

\end{tabular}
}
\end{center}

For the first-order operators, we use the typical definitions
of labelled natural deduction rules \cite{BMV1998}. The only difference lies
in the rules for quantification.  
We may instantiate an universally quantified variable with a horizontally (vertically) 
rigid, if the formula is vertically (horizontally) chop-free. If the formula
is completely chop-free, we may instantiate the variable with an arbitrary term.
Similarly, rigid terms may instantiate \(x\) in arbitrary formulas.
In all cases, a side condition for the instantiation is that \(s\)
respects the sort of \(x\). 

\begin{center}
\scalebox{\factor}{
  \begin{tabular}{cc}
    \AXC{\(\lwff{\Road, V}{\forall x \qsep \phi}\)}
    \AXC{\(\hcf(\phi) \quad \vri(s)\)}
\RL{\(\forall\El\)}
    \BIC{\(\lwff{\Road, V}{\phi\formsubst{x}{s}}\)}
    \DisplayProof
&
    \AXC{\(\lwff{\Road, V}{\forall x \qsep \phi}\)}
    \AXC{\(\vcf(\phi) \quad \hri(s)\)}
\RL{\(\forall\El\)}
    \BIC{\(\lwff{\Road, V}{\phi\formsubst{x}{s}}\)}
    \DisplayProof\\
&\\
  \AXC{\(\lwff{\Road, V}{\forall x \qsep \phi}\)}
    \AXC{\(\hcf(\phi) \quad \vcf(\phi)\)}
\RL{\(\forall\El\)}
    \BIC{\(\lwff{\Road, V}{\phi\formsubst{x}{s}}\)}
    \DisplayProof &
  \AXC{\(\lwff{\Road, V}{\forall x \qsep \phi}\)}
    \AXC{\(\hri(s) \quad \vri(s)\)}
\RL{\(\forall\El\)}
    \BIC{\(\lwff{\Road, V}{\phi\formsubst{x}{s}}\)}
    \DisplayProof \\
  \end{tabular}
}
\end{center}

The elimination and introduction rules for the chop modalities are 
adopted from Rasmussen \cite{Ras2001}, and resemble the rules for
existential quantification. We only show the case for the horizontal
chop, the rules for vertical chopping are obtained straightforwardly,
by replacing horizontal modalities and relations by  the vertical ones.

\begin{center}
\scalebox{\factor}{
\begin{tabular}{cc}
\bottomAlignProof
  \AXC{\(\lwff{\Road, V_1}{\phi}\)}
  \AXC{\(\lwff{\Road, V_2}{\psi}\)}
  \AXC{\(\hreach{V_1}{ V_2}{V}\)}
\RL{\(\chop\In\)}
  \TIC{\(\lwff{\Road, V}{{\phi}\chop{\psi}}\)}
\DisplayProof 
&
\bottomAlignProof
\AXC{\(\lwff{\Road, V}{{\phi}\chop{\psi}}\)}
\AXC{\([\lwff{\Road, V_1}{\phi}]\;[\lwff{\Road, V_2}{\psi}]\; [ \hreach{V_1}{V_2}{V}]\)}
\noLine
\UIC{\(\vdots\)}
\noLine
\UIC{\(\lwff{\Road^\prime, V^\prime}{\chi}\)}
\RL{\(\chop\El\)}
\BIC{\(\lwff{\Road^\prime, V^\prime}{\chi}\)}
\DisplayProof
\end{tabular}
}\end{center}

The chopping of intervals is not ambiguous, i.e., there is  a unique 
view of a certain length at the beginning of a view. This is the 
\emph{single decomposition property} \cite{Dut1995} of interval logics and
captured in the following rules.  
Hence when there are two vertical chops of a view, and the upper parts are of 
equal width, we can derive that the same formulas hold on the lower parts. 
Even though we only show the vertical set of rules, similar
rules hold for the horizontal chopping of views.
\begin{center}
 \scalebox{\factor}{
 \begin{tabular}{c}
    \AXC{\(\lwff{\Road, V_1}{\phi}\)}
    \AXC{\(\lwff{\Road, V_2}{\width = s}\quad\lwff{\Road, V_2^\prime}{\width = s}\)}
    \AXC{\(\vri(s)\quad\vreach{V_1}{V_2}{V}\quad \vreach{V_1^\prime}{V_2^\prime}{V}\)}
\RL{\(VD\)}
    \TIC{\(\lwff{\Road, V_1^\prime}{\phi}\)}
    \DisplayProof\\
\\
    \AXC{\(\lwff{\Road, V_2}{\phi}\)}
    \AXC{\(\lwff{\Road, V_1}{\width = s}\quad\lwff{\Road, V_1^\prime}{\width = s}\)}
    \AXC{\(\vri(s)\quad\vreach{V_1}{V_2}{V}\quad \vreach{V_1^\prime}{V_2^\prime}{V}\)}
\RL{\(VD\)}
    \TIC{\(\lwff{\Road, V_2^\prime}{\phi}\)}
    \DisplayProof\\
\\
  \end{tabular}
}\end{center}

The additivity of length and width can be formalized by the following
rules.
\begin{center} 
\scalebox{\factor}{
\begin{tabular}{c}
  \AXC{\(\lwff{\Road, V_1}{\width = s}\)}
  \AXC{\(\lwff{\Road, V_2}{\width = t}\)}
  \AXC{\(\vri(s) \quad \vri(t) \quad \vreach{V_1}{V_2}{V}\)}
\RL{\(V+\In\)}
  \TIC{\(\lwff{\Road, V}{\width = s +t}\)}
  \DisplayProof\\
\\
\AXC{\(\lwff{\Road, V}{\width = s + t}\)}
\AXC{\(\vri(s) \quad \vri(t)\)}

\AXC{\([\lwff{\Road, V_1}{\width = s}]\;[\lwff{\Road, V_2}{\width = t}]\;[\vreach{V_1}{V_2}{V}]\)}
\noLine
\UIC{\(\vdots\)}
\noLine
\UIC{\(\lwff{\Road^\prime, V^\prime}{\phi}\)}
\RL{\(V+\El\)}
\TIC{\(\lwff{\Road^\prime, V^\prime}{\phi}\)}
\DisplayProof\\
\end{tabular}
}\end{center}

The dynamic modalities are defined along the lines of Basin et al. \cite{BMV1998}. 
If a transition from the current snapshot is possible, the box modalities may
be eliminated and if we can prove that under the assumption of a fresh 
transition \(\alpha\),
\(\phi\) holds on the now reachable snapshot, \(\boxmodal{\alpha} \phi\) holds. 
In the \(\boxmodal{\alpha}\) introduction rule, the label \(\Road^\prime, V^\prime\)
may not occur in any assumption \(\lwff{\Road^\prime, V^\prime}{\phi}\) depends
on, with the exception of \(\Road, V \transition{\alpha} \Road^\prime, V^\prime\). 

\begin{center}
\scalebox{\factor}{
  \begin{tabular}{cc}
\bottomAlignProof
\AXC{\(\Road, V \transition{\alpha} \Road^\prime, V^\prime\)}
    \AXC{\(\lwff{\Road,V}{\boxmodal{\alpha}\phi}\)}
    \RL{\(\boxmodal{\alpha}\El\)}
    \BIC{\(\lwff{\Road^\prime, V^\prime}{ \phi}\)}
    \DisplayProof
&
\bottomAlignProof
\AXC{\([\Road, V \transition{\alpha} \Road^\prime, V^\prime]\)}
\noLine
\UIC{\(\vdots\)}
\noLine
\UIC{\(\lwff{\Road^\prime, V^\prime}{\phi}\)}
\RL{\(\boxmodal{\alpha}\In\)}
\UIC{\(\lwff{\Road,V}{\boxmodal{\alpha}\phi}\)}
\DisplayProof\\
  \end{tabular}
}\end{center}

Finally, we have to define how the spatial atoms behave with respect to
occurring transitions. There are two types of rules in general, 
\emph{stability rules} and \emph{activity rules}. Stability rules define
which atoms stay true after a snapshot changes according to a
certain transition. The truth of all reservation and claims of cars not involved
in the transition are unchanged. Only one stability rule for creating
reservations includes the car which is the source of the transition.  
We will show this rule and one example for typical stability.
 The \emph{activity rules}  state how the reservations and claims
 of cars will change according to the transitions.

The following stability rules show that whenever a car creates a new
claim, the reservations and claims of other cars are unchanged. 
We have similar stability rules for the other types of transitions.
\begin{center}
\scalebox{\factor}{
  \begin{tabular}{c}
    \AXC{\(\lwff{\Road, V}{\claimed{c}}\)}
    \AXC{\(\Road, V \transition{\lndclaim{d}} \Road^\prime, V\)}
    \AXC{\(\lwff{\Road, V}{c \neq d}\)}
    \RL{\(\transition{\lndclaim{c}}\Stab\)}
    \TIC{\(\lwff{\Road^\prime, V}{\claimed{c}}\)}
    \DisplayProof \\\\
    \AXC{\(\lwff{\Road, V}{\reserved{c}}\)}
    \AXC{\(\Road, V \transition{\lndclaim{d}} \Road^\prime, V\)}
    \AXC{\(\lwff{\Road, V}{c \neq d}\)}
    \RL{\(\transition{\lndclaim{c}}\Stab\)}
    \TIC{\(\lwff{\Road^\prime, V}{\reserved{c}}\)}
    \DisplayProof 
  \end{tabular}
  }
\end{center}

The
activity rule for \(\lndclaim{c}\) implies two properties.
First, a claim may only be created when only one reservation exists. 
Second, the newly created claim resides on one side of the existing
reservation. Observe that we require the view under consideration to comprise
both adjacent lanes of the reservation. If we dropped this assumption (i.e., 
removed the subformulas \(\width = 1\)), it would be possible for
the newly created claim to reside outside of the view \(V\), and hence
the conclusion would not be satisfied.
\begin{center}
\scalebox{\factor}{
  \begin{tabular}{c}
    \AXC{\(\lwff{\Road, V}{\vchop{\lnot(\reserved{c} \lor \claimed{c}) \land \width = 1}{\vchop{\reserved{c}}{\lnot(\reserved{c}\lor \claimed{c})\land \width = 1}}}\)}
    \AXC{\(\Road, V \transition{\lndclaim{d}} \Road^\prime, V\)}
    \AXC{\(\lwff{\Road, V}{c = d}\)} 
    \RL{\(\transition{\lndclaim{c}}\Act\)}
    \TIC{\(\lwff{\Road^\prime, V}{\vchop{\lnot(\reserved{c} \lor \claimed{c}) \land\width = 1}{\vchop{\reserved{c}}{\claimed{c}}} \lor \vchop{\vchop{\claimed{c}}{\reserved{c}}}{\lnot(\reserved{c} \lor \claimed{c}) \land\width = 1} }\)}
    \DisplayProof 
\\
  \end{tabular}
}\end{center}
Activity rules for the creation of reservations in between traffic
snapshots are: 
\begin{center}
\scalebox{\factor}{
  \begin{tabular}{c}
    \AXC{\(\lwff{\Road, V}{\claimed{c}}\)}
    \AXC{\(\Road, V \transition{\lndres{d}} \Road^\prime, V\)}
    \AXC{\(\lwff{\Road, V}{c = d}\)}
    \RL{\(\transition{\lndres{c}}\Act_1\)}
    \TIC{\(\lwff{\Road^\prime, V}{\reserved{c}}\)}
    \DisplayProof
\\\\ 
    \AXC{\(\lwff{\Road, V}{\reserved{c}}\)}
    \AXC{\(\Road, V \transition{\lndres{d}} \Road^\prime, V\)}
    \AXC{\(\lwff{\Road, V}{c = d}\)}
    \RL{\(\transition{\lndres{c}}\Act_2\)}
    \TIC{\(\lwff{\Road^\prime, V}{\reserved{c}}\)}
    \DisplayProof
\end{tabular}
}\end{center}
The following
activity rules define
 the withdrawal of reservations and claims. 
\begin{center}
\scalebox{\factor}{
   \begin{tabular}{cc}
    \AXC{\(\lwff{\Road, V}{\vchop{\reserved{c}}{\reserved{c}}}\)}
    \AXC{\(\Road, V \transition{\lndwdres{d}} \Road^\prime, V\)}
    \AXC{\(\lwff{\Road, V}{c = d}\)}
    \RL{\(\transition{\lndwdres{c}}\Act\)}
    \TIC{\(\lwff{\Road^\prime, V}{\vchop{\reserved{c}}{\lnot\reserved{c}} \lor \vchop{\lnot\reserved{c}}{\reserved{c}} }\)}
    \DisplayProof
&
    \AXC{\(\Road, V \transition{\lndwdclaim{c}} \Road^\prime, V\)}
    \RL{\(\transition{\lndwdclaim{c}}\Act\)}
    \UIC{\(\lwff{\Road^\prime, V}{\lnot\claimed{c}}\)}
    \DisplayProof
  \end{tabular}
}\end{center}

Note that we cannot define rules relating the spatial situations 
along evolutions of time. This is due to the fact that we lost all 
knowledge about the concrete dynamics of the underlying
semantics. Hence all constraints of the cars' behaviour 
have to be explicitly 
defined within EMLSL, like the exemplary requirement for 
a safe distance controller  in Sect.\ref{sec:mlsl}.

We also have rules for ``backwards'' reasoning, i.e., if our current
snapshot is reachable from another, 
we may draw conclusions about the originating snapshot. Again,
we differentiate between activity and stability rules (omitted here).
\begin{center}
\scalebox{\factor}{
  \begin{tabular}{c}
    \AXC{\(\lwff{\Road^\prime, V}{\reserved{c}}\)}
    \AXC{\(\Road, V \transition{\lndres{d}} \Road^\prime, V\)}
    \AXC{\(\lwff{\Road, V}{c = d}\)}
    \RL{\(\backtrans{\lndres{c}}\Act\)}
    \TIC{\(\lwff{\Road, V}{\reserved{c} \lor \claimed{c}}\)}
    \DisplayProof
  \end{tabular}
}\end{center}

\begin{center}
\scalebox{\factor}{
  \begin{tabular}{c}
    \AXC{\(\lwff{\Road^\prime, V}{\claimed{c}}\)}
    \AXC{\(\Road, V \transition{\lndclaim{d}} \Road^\prime, V\)}
    \AXC{\(\lwff{\Road, V}{c = d}\)}
    \RL{\(\backtrans{\lndclaim{c}}\Act\)}
    \TIC{\(\lwff{\Road, V}{\lnot \claimed{c}}\)}
    \DisplayProof
  \end{tabular}
}\end{center}

\begin{center}
\scalebox{\factor}{
  \begin{tabular}{c}
    \AXC{\(\lwff{\Road^\prime, V}{\vchop{\width = 1}{\vchop{\reserved{c}}{\width = 1}}}\)}
    \AXC{\(\Road, V \transition{\lndwdres{d}} \Road^\prime, V\)}
    \AXC{\(\lwff{\Road, V}{c = d}\)}
    \RL{\(\backtrans{\lndwdres{c}}\Act\)}
    \TIC{\(\lwff{\Road, V}{\vchop{\reserved{c}}{\vchop{\reserved{c}}{\width=1}} \lor \vchop{\width=1}{\vchop{\reserved{c}}{\reserved{c}}} }\)}
    \DisplayProof
  \end{tabular}
}\end{center}

Observe that we can not reason backwards along withdrawals of claims,
since these may  
be taken anytime, even when no claim previously existed 
(cf. Def.~\ref{def:transitions}).

\begin{thm}
\label{thm:soundness}
 The calculus of labelled natural deduction for EMLSL is sound. 
\end{thm}

\proof
Since we do not have any inference rules for the transitions between snapshots and
the rules for relations between views are straightforward, we only have to 
consider the case of derivations of labelled formulas. 

We proceed by induction on the length of derivations to show 
\(\Gamma, \Delta \proofs \lwff{\Road, V}{\phi}\) implies 
\(\Gamma, \Delta \models \lwff{\Road, V}{\phi}\). 

If \(\lwff{\Road, V}{\phi} \in \Gamma\), then trivially \(\Gamma, \Delta \models \lwff{\Road, V}{\phi}\).

For the induction step, assume that for all smaller derivations \(\Gamma_i, \Delta_i \proofs \lwff{\Road_i, V_i}{\phi_i}\), we already have \(\Gamma_i, \Delta_i \models \lwff{\Road_i, V_i}{\phi_i}\).

We only show some exemplary cases, for the rigidity rules, the elimination
of the universal quantifier. Proofs for the other
rules are either analogous, or can be straightforwardly infered from
the work of Rasmussen \cite{Ras2001}, Basin et al. \cite{BMV1998} and 
Vigan\`{o} \cite{Vig2000}. However, we explicitly prove the soundness
of all activity rules for reasoning forwards and backwards along
traffic transitions.

The last step in the derivation is an application of \(R_H\). Then \(\phi\) is 
horizontally rigid. Let 
\(\Gamma, \Delta \proofs \lwff{\Road, V}{\phi}\), 
\(\Delta^\prime = \{\hreach{V_1}{V_2}{V}\} \cup \Delta\) 
and \(\val \models (\Gamma, \Delta^\prime)\) and hence also
\(\val \models (\Gamma, \Delta)\). By the induction hypothesis, we have
\(\val \models \lwff{\Road, V}{\phi}\). By Def.~\ref{def:lwff_sat}, we get
\(\Road, V, \val \models\phi\).
Now we may use Lemma~\ref{lem:rigidity} twice,
once to get  
\(\Road^\prime, V, \val \models \phi\) (since \(\phi\) is also dynamically rigid) 
and the second time to
get \(\Road^\prime, V_1, \val \models \phi\). Finally, we have
\(\val \models \lwff{\Road^\prime, V_1}{\phi}\). The other cases of this
rule are similarly proven.

The last step in the derivation is an application of the first variant
of \(\forall \El\). 
 Then \(\phi\) is horizontal chop free and \(s\) is 
vertically rigid. 
Let \(\Gamma, \Delta \proofs \lwff{\Road,V}{\forall x \qsep \phi}\) and
\(\val \models (\Gamma,\Delta)\). By the induction hypothesis,
 we have \(\val \models \lwff{\Road,V}{\forall x \qsep \phi}\), i.e.,
\(\Road,V,\val \models \forall x \qsep \phi\). Since \(\phi\) is horizontal
chop free, it may at most contain a vertical chop. However, the value 
\(\val_V(s)\) is constant on all vertical subviews of \(V\), hence also for
all subformulas of \(\phi\). All in all, \( \Road,V,\val \models \phi\formsubst{x}{s}\),
i.e., \(\val \models \lwff{\Road,V}{\phi\formsubst{x}{s}}\). The other variants
of the quantifier elimination are analogous.

The last step in the derivation is an application of \(\transition{\lndwdres{c}}\Act\).
Then let 
\begin{align*}
\Gamma_1, \Delta\proofs \lwff{\Road, V}{\vchop{\reserved{c}}{\reserved{c}}} \text{ and }\Gamma_2, \Delta\proofs \lwff{\Road, V}{c = d},
\end{align*}  
 with  
\(\Gamma_1 \cup \Gamma_2 = \Gamma\)  
and \(\Road,V \transition{\lndwdres{d}} \Road^\prime,V \in  \Delta\), which
by the induction hypothesis implies both  
\begin{align*}
\Gamma_1, \Delta\models \lwff{\Road, V}{\vchop{\reserved{c}}{\reserved{c}}}  \text{ and } 
\Gamma_2, \Delta\models \lwff{\Road, V}{c = d}.
\end{align*}
We assume \(\val \models (\Gamma, \Delta)\), i.e., \(\val \models (\Gamma_1, \Delta)\)
and \(\val \models (\Gamma_2, \Delta)\). Hence 
\begin{align*}
\val \models \lwff{\Road, V}{\vchop{\reserved{c}}{\reserved{c}}}, \, 
\val \models \lwff{\Road, V}{c = d} \text{ and } 
\val \models \Road,V \transition{\lndwdres{d}} \Road^\prime,V.
\end{align*}
Let \(\vreach{V_1}{V_2}{V}\), such that 
\(\Road, V_1, \val \models \reserved{c}\) and 
\(\Road, V_2, \val \models \reserved{c}\) with \(V_i = (L_i, X_i, E)\). 
We know that there is a \(n_0\), such that 
\( \Road \transition{\stopmoveToLane{\val(d)}{n_0}} \Road^\prime\).
Let \(n_0 \in L_1\). Then by Definition~\ref{def:transitions}, we have 
that \(\laneReservation^\prime_{V_2}(\val(d)) = \emptyset\), which means
\(\Road^\prime, V_2, \val \not\models \reserved{c}\), that is,
\(\Road^\prime, V_2, \val \models \lnot \reserved{c}\). Furthermore, we have 
\(n_0 \in \laneReservation^\prime_{V_2}(\val(d))\), i.e.,
\(\Road^\prime, V_1, \val \models \reserved{c}\). By definition of the vertical chop, we
get
\begin{align*}
\Road^\prime, V, \val \models \vchop{\reserved{c}}{\lnot \reserved{c}}
\end{align*}
 and  hence 
\begin{align*}
\Road^\prime, V, \val \models \vchop{\reserved{c}}{\lnot \reserved{c}} \lor \vchop{\lnot\reserved{c}}{ \reserved{c}}.
\end{align*} 
If \(n_0 \in L_2\), the reasoning is analogous. All in all, we get that 
\begin{align*}
\val \models \lwff{\Road^\prime, V}{\vchop{\reserved{c}}{\lnot \reserved{c}} \lor \vchop{\lnot\reserved{c}}{ \reserved{c}}}.
\end{align*}

The last step in the derivation is an application of \(\transition{\lndres{c}}\Act\).
Then let 
\(\Gamma_1, \Delta\proofs \lwff{\Road, V}{\claimed{c}}\),  
\(\Gamma_2, \Delta\proofs \lwff{\Road, V}{c = d}\), with  
\(\Gamma_1 \cup \Gamma_2 = \Gamma\) and
\(\Road, V \transition{\lndres{d}} \Road^\prime, V \in \Delta\). 
By the
induction hypothesis we get
\(\Gamma_1, \Delta\models \lwff{\Road, V}{\claimed{c}}\),  
\(\Gamma_2, \Delta\models \lwff{\Road, V}{c = d}\). 
Now assume \(\val \models (\Gamma, \Delta)\). That is, 
\(\val(c) = \val(d)\) and \(\Road, V, \val \models \claimed{c}\) and
\(\Road \transition{\moveToLane{\val(d)}} \Road^\prime\). Thus 
\(\laneClaim_V(\val(c)) = \ViewLanes\), where \(\ViewLanes\) are the lanes
of \(V\).
By Definition~\ref{def:transitions} we get that 
\(\laneReservation^\prime(\val(d)) = \laneReservation(\val(d)) \cup \laneClaim(\val(d))\)
and, since \(\val(c) = \val(d)\), 
\(\laneReservation^\prime_V(\val(c)) = \laneClaim_V(\val(c)) = \ViewLanes\).
So \(\Road^\prime, V, \val \models \reserved{c}\) and by that 
\(\val \models \lwff{\Road^\prime, V}{\reserved{c}}\).


Let the last step of the derivation be an application of 
\(\transition{\lndwdclaim{c}}\Act\). Furthermore, let 
\(\Gamma, \Delta \proofs \Road, V \transition{\lndwdclaim{c}} \Road^\prime,V\),
i.e. \( \Road, V \transition{\lndwdclaim{c}} \Road^\prime,V \in \Delta\). Hence
\(\Road \transition{\stopclaimLane{\val(c)}} \Road^\prime\) is true. That is,
\(\laneClaim^\prime(\val(c)) = \emptyset\) which implies 
\(\Road^\prime, V, \val \models \lnot \claimed{c}\). Thus 
\(\val \models \lwff{\Road^\prime, V}{\lnot \claimed{c}}\).


Let the last step of the derivation be an application of 
\(\transition{\lndclaim{c}}\Act\). Furthermore we assume 
\begin{align*}
\Gamma_1, \Delta \proofs \lwff{\Road, V}{\vchop{\lnot(\reserved{c} \lor \claimed{c}) \land \width = 1}{\vchop{\reserved{c}}{\lnot(\reserved{c}\lor \claimed{c})\land \width = 1}}},
\end{align*}
\(\Gamma_2, \Delta \proofs \lwff{\Road,V}{c = d}\)
and \(\Road, V \transition{\lndclaim{d}} \Road^\prime, V \in \Delta\). Furthermore,
let \(\Gamma = \Gamma_1 \cup \Gamma_2\) and \(\val \models (\Gamma, \Delta)\).
That is, for \(V = (\ViewLanes, \Extension, E)\), we know that 
\(\ViewLanes\) contains exactly three elements, say \(\ViewLanes = \{n_1, n_2, n_3\}\)
and that \(\laneReservation(\val(c)) = \{n_2\}\) and \(\laneClaim(\val(c)) = \emptyset\).
Now consider \(\Road^\prime\). Since \(\val \models \Road,V \transition{\lndclaim{d}} \Road^\prime, V\), we have \(\Road \transition{\claimLane{\val(d)}{n^\prime}} \Road^\prime\) for
either \(n^\prime = n_1\) or \(n^\prime = n_3\). Furthermore, due to \(\val(c) = \val(d)\)
we know that \(\laneClaim^\prime(\val(c)) = \{n^\prime\}\). Say \(n^\prime = n_1\). Then
\(\Road^\prime, V^{\{n_1\}}, \val \models \claimed{c}\). Note that the
extension of \(V^{\{n_1\}}\) has to be greater than zero, since the subview
\(V^{\{n_2\}}\) already satisfies \(\reserved{c}\).
Due to 
\(\laneReservation = \laneReservation^\prime\), we get
\begin{align*}
&&\Road^\prime, V \val &\models   
\vchop{\lnot(\reserved{c} \lor \claimed{c}) \land \width = 1}{\vchop{\reserved{c}}{\claimed{c}}}\\
\Rightarrow&& \val &\models   
\lwff{\Road^\prime, V}{\vchop{\lnot(\reserved{c} \lor \claimed{c}) \land \width = 1}{\vchop{\reserved{c}}{\claimed{c}}}}\\
\Rightarrow&& \val &\models   
\lwff{\Road^\prime, V}{\vchop{\lnot(\reserved{c} \lor \claimed{c}) \land \width = 1}{\vchop{\reserved{c}}{\claimed{c}}} \lor
\vchop{\claimed{c}}{\vchop{\reserved{c}}{\lnot(\reserved{c} \lor \claimed{c}) \land \width = 1}}
 }.
\end{align*}
The case where \(n^\prime = n_3\) is similar.

The last step of the derivation is an applicaton of \(\backtrans{\lndclaim{c}}\Act\). 
Then \(\Gamma_1, \Delta\proofs \lwff{\Road^\prime, V}{\claimed{c}}\),  
\(\Gamma_2, \Delta\proofs \lwff{\Road, V}{c = d}\), with  
\(\Gamma_1 \cup \Gamma_2 = \Gamma\)  
and \(\Road,V \transition{\lndwdres{d}} \Road^\prime,V \in  \Delta\), which
by the induction hypothesis implies both  
\(\Gamma_1, \Delta\models \lwff{\Road^\prime, V}{\claimed{c}}\) and  
\(\Gamma_2, \Delta\models \lwff{\Road, V}{c = d}\). We assume \(\val \models (\Gamma_1,\Delta)\) and \(\val \models (\Gamma_2,\Delta)\). Since \(c = d \) is dynamically rigid,
we also have \(\val \models \lwff{\Road^\prime, V}{c = d}\) by Lemma~\ref{lem:rigidity}.
So \(\val_V(c) = \val_V(d)\). There can only be a transition creating
a new claim for \(\val_V(c)\) from \(\Road\) to \(\Road^\prime\), if 
\(\laneClaim(\val_V(c)) = \emptyset\)
on \(\Road\). Hence, for each view \(V^\prime\), 
\(\Road, V^\prime,\val \models \lnot\claimed{c}\). Hence in particular
 \(\Road, V,\val \models \lnot\claimed{c}\), i.e.,  
\(\val \models \lwff{\Road, V}{\lnot\claimed{c}}\).

Let the last step of the derivation be an application of
\(\backtrans{\lndwdres{c}}\Act\) and
let 
\begin{align*}
\Gamma_1, \Delta \proofs \lwff{\Road^\prime, V}{\vchop{\width=1}{\vchop{\reserved{c}}{\width=1}}},\,
\Gamma_2, \Delta \proofs \lwff{\Road^\prime, V}{c = d} \text{ and } 
\Road, V \transition{\lndwdres{d}} \Road^\prime\in\Delta. 
\end{align*}
Now assume
\(\val \models (\Gamma_1 \cup \Gamma_2, \Delta)\). By the induction hypothesis, 
we get 
\begin{align*}
\val \models \lwff{\Road^\prime, V}{\vchop{\width=1}{\vchop{\reserved{c}}{\width=1}}}
\text{ and } \val \models  \lwff{\Road^\prime, V}{c = d}. 
\end{align*}
By that we know that
the set of lanes \(\ViewLanes\) of \(V =(\ViewLanes, \Extension, E)\) contains
exactly three elements, say \(\ViewLanes = \{n_1, n_2, n_3\}\) and by the
semantics of the transitions (see Def.~\ref{def:transitions}) and EMLSL
(see Def.~\ref{def:semantics-MLSL}), we get
\(\laneReservation^\prime(\val(c)) = \{n_2\}\). The transition exists only, 
when \(|\laneReservation(\val(c))| = 2\) and \(n_2 \in \laneReservation(\val(c))\),
so there are only two possibilities (due to the sanity conditions of
Def.~\ref{def:snapshots}): \(n_1 \in \laneReservation(\val(c))\) or 
\(n_3 \in \laneReservation(\val(c))\). Say \(n_1 \in \laneReservation(\val(c))\).
Then 
\begin{align*}
\Road, V, \val \models \vchop{\width = 1}{\vchop{\reserved{c}}{\reserved{c}}}
\end{align*}
and hence 
\begin{align*}
\val \models \lwff{\Road, V}{\vchop{\reserved{c}}{\vchop{\reserved{c}}{ \width=1}}\lor \vchop{\width = 1}{\vchop{\reserved{c}}{\reserved{c}}}   }.
\end{align*} 
The case for \(n_3 \in \laneReservation(\val(c))\) is similar. 

Let the last step in the derivation be an application of 
\(\backtrans{\lndres{c}} \Act\) and let furthermore 
\(\Gamma_1, \Delta \proofs \lwff{\Road^\prime, V}{\reserved{c}}\), 
\(\Gamma_2, \Delta \proofs \lwff{\Road, V}{ c = d}\) and 
\(\Road, V \transition{\lndres{d}} \Road^\prime, V \in \Delta\). 
By the induction hypothesis, we get 
\(\Gamma_1, \Delta \models \lwff{\Road^\prime, V}{\reserved{c}}\) and 
\(\Gamma_2, \Delta \models \lwff{\Road, V}{ c = d}\). 
Now let 
\(\Gamma = \Gamma_1 \cup \Gamma_2\) and \(\val \models (\Gamma, \Delta)\).
We then know that \(\laneReservation^\prime_V(c) = \{n\}\) where 
\(V = (\ViewLanes, \Extension, E)\) with \(\ViewLanes = \{n\}\) and \(\|X\| > 0\). 
By Def.~\ref{def:transitions}  and \(\val(c) = \val(d)\) we get that  
\(\laneReservation^\prime(\val(c)) = \laneReservation(\val(c)) \cup \laneClaim(\val(c))\). 
If \(n \in \laneReservation(\val(c))\), we have \(\Road, V,\val \models \reserved{c}\),
which implies \(\Road, V, \val \models \reserved{c} \lor \claimed{c}\). 
Similarly, if \(n \in \laneClaim(\val(c))\), we get \(\Road, V,\val \models \claimed{c}\),
which implies \(\Road, V,\val \models\reserved{c} \lor \claimed{c}\).
That is, \(\val \models \lwff{\Road, V}{\reserved{c} \lor \claimed{c}}\).
\qed

Since models of EMLSL are based on the real numbers, we cannot hope for
a complete deduction system. Even if we used an infinite and dense field
instead of the real numbers, it is in no way obvious, whether the
resulting proof system would be complete. Typical approaches for
constructing maximally consistent  sets \cite{Vig2000} 
are not directly applicable, since 
they may result in an infinite number of lanes in the canonical model.

As an example, we derive a variant
of the \emph{reservation lemma}, which we proved informally in our previous 
work \cite{HLOR2011}. 
\begin{lem}[Reservation]
\label{lem:reservation} A reservation of a car \(c\) observed 
directly after \(c\) created a reservation, 
was either already present or 
is due to a previously  existing 
claim. I.e., assuming \(\Road, V \transition{\lndres{c}} \Road^\prime, V\), the
 formula 
\(
  (\reserved{c} \lor \claimed{c}) \equivalent \boxmodal{\lndres{c}} \reserved{c}
\)
holds.
Hence  
\begin{align*}
 \{\Road, V \transition{\lndres{c}} \Road^\prime, V\} \proofs \lwff{\Road, V}{(\reserved{c} \lor \claimed{c}) \equivalent \boxmodal{\lndres{c}} \reserved{c}}.
\end{align*}
\end{lem}
\proof
The existence of the transition is of major importance 
for the elimination
of the box modality in the proof using the backwards reasoning rule. 
For reasons of simplicity, we use a variant of the stability rules and activity rules,
where  \(d\) in the transition has been replaced by \(c\), and hence we do
not need the extra assumption of \(\lwff{\Road, V}{c = d}\). 
We use two auxiliary derivations \(\Pi_\Stab\) and \(\Pi_\Act\), which 
allow us to infer the existence of a 
reservation on the snapshot after taking a transition.

\begin{center}
\(\Pi_\Stab\):
\scalebox{.85}{
  \AXC{\([\lwff{\Road,V}{\reserved{c}}]_1\)}
  \AXC{\([\Road, V \transition{\lndres{c}} \Road^\prime, V]_2\)}
  \BIC{\(\lwff{\Road^\prime,V}{\reserved{c}}\)}
\DisplayProof
}

\(\Pi_\Act\):
\scalebox{.85}{
  \AXC{\([\lwff{\Road,V}{\claimed{c}}]_1\)}
  \AXC{\([\Road, V \transition{\lndres{c}} \Road^\prime, V]_2\)}
  \BIC{\(\lwff{\Road^\prime,V}{\reserved{c}}\)}
\DisplayProof
}
\end{center}

Derivation of \(\proofs \lwff{\Road,V}{(\reserved{c}\lor\claimed{c}) \rightarrow \boxmodal{\lndres{c}} \reserved{c}}\).
\begin{center}
\scalebox{.85}{
\AXC{\(\Pi_\Stab\)}

\AXC{\(\Pi_\Act\)}
  
  \AXC{\([\lwff{\Road, V}{\reserved{c}\lor\claimed{c}}]_3\)}
  \LL{\(\lor \El_1\)}
  \TIC{\(\lwff{\Road^\prime,V}{\reserved{c}}\)}
  \RL{\(\boxmodal{\lndres{c}}\In_2\)}
  \UIC{\(\lwff{\Road,V}{\boxmodal{\lndres{c}} \reserved{c}}\)}
 \RL{\(\rightarrow \In_3\)}
  \UIC{\(\lwff{\Road,V}{(\reserved{c}\lor\claimed{c}) \rightarrow \boxmodal{\lndres{c}} \reserved{c}}\)}
 \DisplayProof
}  
\end{center}

Derivation of \(\{\Road, V \transition{\lndres{c}} \Road^\prime, V\} \proofs \lwff{\Road,V}{\boxmodal{\lndres{c}} \reserved{c} \rightarrow (\reserved{c}\lor\claimed{c})}\).

\begin{center}
\scalebox{.85}{
  \AXC{\([\lwff{\Road,V}{\boxmodal{\lndres{c}}\reserved{c}}]_1\)}
  \AXC{\(\Road, V \transition{\lndres{c}} \Road^\prime, V\)}
  \RL{\(\boxmodal{\lndres{c}}\El\)}
  \BIC{\(\lwff{\Road^\prime,V}{\reserved{c}}\)}
  \AXC{\(\Road, V \transition{\lndres{c}} \Road^\prime, V\)}
  \RL{\(\backtrans{\lndres{c}}\)}
  \BIC{\(\lwff{\Road,V}{\reserved{c}\lor\claimed{c}}\)}
  \RL{\(\rightarrow \In_1\)}
  \UIC{\(\lwff{\Road,V}{\boxmodal{\lndres{c}}\reserved{c} \rightarrow (\reserved{c}\lor\claimed{c})}\)} 
  \DisplayProof
}
\end{center}\qed

A second example showing how the rigidity rules and chopping rules interact is 
the following.

\begin{lem}[Independence of Length and Width]
\label{lem:length_and_width}
For all traffic snapshots and views, the length of the view is the same
on all vertical subviews, i.e.
\begin{align*}
  \lwff{\Road, V}{ \vchop{\length = x}{\length = x} \equivalent \length = x}.
\end{align*}
\end{lem}
\proof
First we show \(\proofs \lwff{\Road, V}{\vchop{\length = x}{\length = x} \implies \length = x}\).
We define two auxiliary subderivations, which are in essence applications of the 
rules for rigidity.

\begin{center}
\(\Pi^R_i\):
\scalebox{.85}{
\AXC{}
\UIC{\(\vri(\length)\)}
\AXC{}
\UIC{\(\vri(x)\)}
\BIC{\(\vri(\length = x)\)}
\AXC{\([\vreach{V_1}{V_2}{V}]_1\)}
\AXC{\([\lwff{\Road, V_i}{\length = x}]_1\)}
\LL{\(R_V\)}
\TIC{\(\lwff{\Road, V}{\length = x}\)}
\DisplayProof
}
\end{center}

The eliminations of assumptions are due to  the chop-elimination in the 
following proof.

\begin{center}
\scalebox{.85}{
\AXC{\(\left[\lwff{\Road, V}{\vchop{\length =x }{\length = x}}\right]_2\)}

 \AXC{\(\Pi^R_1\)}

 \AXC{\(\Pi^R_2\)}
\RL{\(vC\El_1\)}
\TIC{\(\lwff{\Road, V}{\length = x}\)}
\RL{\(\implies\In_2\)}
\UIC{\(\lwff{\Road, V}{\vchop{\length =x }{\length = x} \implies \length = x}\)}
\DisplayProof
}
\end{center}

Now we turn to the other direction. Here, we need to assume that the decomposition
of the view into two subviews is possible, i.e., the derivation contains an
application of the elimination rule for the existential quantifier for views. 

For reasons of readability, we define two subderivations \(\Pi^V_i\). In each
of these derivations, we infer from the assumption, that \(V\) has an extension
of length \(x\), that also the subview \(V_i\) has an extension of length \(x\). 

\begin{center}
\(\Pi^V_i\):
\scalebox{.85}{
\AXC{}
\UIC{\(\vri(\length)\)}
\AXC{}
\UIC{\(\vri(x)\)}
\BIC{\(\vri(\length = x)\)}
\AXC{\([\vreach{V_1}{V_2}{V}]_2\)}
\AXC{\([\lwff{\Road, V}{\length = x}]_1\)}
\RL{\(R_V\)}
\TIC{\(\lwff{\Road, V_i}{\length = x}\)}
\DisplayProof
}
\end{center}

The eliminations of the assumptions indicated by the indices 
are due to the rules used in the
final derivation, as follows. 

\begin{center}
\scalebox{.85}{
\AXC{}
\UIC{\(\vexists V^\prime, V^{\prime\prime}(\vreach{V^\prime}{V^{\prime\prime}}{V})\)}
\AXC{\(\Pi^V_1\)}

\AXC{\(\Pi^V_2\)}

\AXC{\([\vreach{V_1}{V_2}{V}]_2\)}
\RL{\(vC\In\)}
\TIC{\(\lwff{\Road, V}{\vchop{\length = x}{\length = x}}\)}

\RL{\(\vexists\El_2\)}
\BIC{\(\lwff{\Road, V}{\vchop{\length = x}{\length = x}}\)}
\RL{\(\implies\In_1\)}
\UIC{\(\lwff{\Road, V}{\length = x \implies \vchop{\length = x}{\length = x}}\)}
\DisplayProof
}
\end{center}
By the combination of both these derivations and the usual shortcut
for biimplication introduction, we get the desired result. \qed


\section{Related and Future Work}
\label{sec:conc}

Most related work on spatial logics is focused on purely qualitative 
spatial reasoning 
\cite{vBB2007},
e.g., the expressible properties  concern topological relations 
\cite{Randell1992}. Logics expressing quantitative spatial properties
are rare, an example is Sch\"afer's Shape Calculus (SC)~\cite{Sch2005}, 
which is a very general extension of DC. Contrasting SC, the focus of EMLSL
lies on a restricted field of application, i.e., highway traffic.  

 EMLSL  is an instance of a multi-dimensional and 
multi-modal logic \cite{GKWZ2003}, since it consists of various
different modal operators, which are not interdefinable. However,
the modalities are strongly interconnected, e.g. the creation 
of a reservation only has  an effect, if there was a preceeding
creation of a claim for the same car. Hence EMLSL is not simply 
a fusion of the corresponding uni-modal languages, but presumably
determined by a class of suitable  product frames. It is worthwile to
study, which properties the parts of these frames are required to have.

Labelled natural deduction for (multi-)modal logics has been
studied intensely recently. 
E.g.,  when the rules for relational formulas can be defined
with  horn  clauses as antecedents, nice meta-theoretical
properties like normalization of proofs can be established~\cite{BMV1998,Vig2000}.
In intuitionistic modal logic, similar results are obtained, 
when the relational theory is defined using only
geometric sequents \cite{Sim1994}. Unfortunately, even with our restricted set of rules for view relations,
these results do not carry over to our setting, since 
we made use of existential quantification on views.
Consider, e.g., the proof of Lemma~\ref{lem:length_and_width}.
There the relational rule for the elimination of existential
quantification over views is used within an otherwise purely logical deduction.
Still we would like to explore how
rules for the manipulation of traffic snapshots could 
blend in. However, due to the complex
internal structure of traffic snapshots, we do not expect such 
rules to be definable by horn clauses. 

The labelling algebra  is deeply intertwined
with the predicates and operators of EMLSL. Changes in the former
would induce adaptations in the latter and vice versa. For example, a 
possible
extension would be to exchange the dynamical modality \(\boxmodal{\tau}\)
by a metric variant \(\boxmodal{[a,b]}\), where \(a\) and \(b\) are
elements of a suitable domain of time, say \(\R\). This change
would have to be reflected in the labelling algebra by replacing
the relation \(\transition{\tau}\) with transitions labelled by
real numbers (or real-valued variables). Then rules expressing
the properties of these relational formulas may be added, e.g., for
additivity of durations.

Rasga et al.  investigated the fibring \cite{Caleiro2005} of
labelled deductive systems \cite{RSSV2002}. We assume that the
deduction system of Sec.~\ref{sec:lnd} is an instance of such
a fibring, where the Boolean operators are shared between all
deduction systems involved. A further 
classification of EMLSL (or a suitable subset)
and its proof system within the framework of fibring and 
multi-dimensional logics would be of interest in order to use preservation
results concerning, e.g., decidability. 

To further increase the possible applications of EMLSL, we seek
 to introduce a global box modality \(\Box\). Intuitively, 
a formula \(\Box\phi\) shall express that \(\phi\) is an invariant
over all possible sequences of transitions. This modality is not 
expressible with the 
help of the other modalities and is intuitively similar to 
an iteration of the transitions like in dynamic logic \cite{HTD2000}.
Finally, an implementation  within
a general theorem prover like Isabelle~\cite{Pau1994} similar
to implementations for modal or interval logics~\cite{BMV1998,Vig2000,Ras2001}  would
increase the usefulness of the proof system.   

\bibliographystyle{alpha}
\bibliography{lit}

\end{document}